\documentclass[a4paper,preprint,10pt,3p]{elsarticle}

\RequirePackage{mathtools}  
\RequirePackage{amssymb}    
\RequirePackage{siunitx}    

\RequirePackage{tabularx}   
\RequirePackage{booktabs}   
\RequirePackage{longtable}  
\RequirePackage{multirow}   

\RequirePackage{graphicx}   
\RequirePackage{float}      
\RequirePackage[labelfont=bf,justification=centering,footnotesize]{caption} 
\RequirePackage{subcaption} 
\RequirePackage{pdfpages}   

\RequirePackage{xcolor}     
\RequirePackage{tikz}       
\RequirePackage{xspace}     
\RequirePackage{microtype}  

\RequirePackage{geometry}   
\RequirePackage{titlesec}   
\RequirePackage{titletoc}   
\RequirePackage{fancyhdr}   
\RequirePackage{enumitem}   
\RequirePackage{etoolbox}   
\RequirePackage{iftex}      
\RequirePackage{datetime}   
\RequirePackage{hyperref} 
\RequirePackage[noabbrev,nameinlink]{cleveref} 

\RequirePackage{listings} 

\usepackage{amsmath}

\usepackage{algorithm}
\usepackage[noend]{algpseudocode}
\usepackage{comment}
\usepackage{placeins}

\DeclareMathOperator*{\argmin}{arg\,min}

\usepackage{array}
\newcolumntype{L}[1]{>{\raggedright\let\newline\\\arraybackslash\hspace{0pt}}m{#1}}

\newdefinition{rmk}{Remark}

\begin{document}
\begin{frontmatter}

\title{Structure-Preserving Hyper-Reduction and Temporal Localization for Reduced Order Models of Incompressible Flows}
\author[1,2]{R.B. Klein\corref{cor1}}
\ead{rbk@cwi.nl}
\author[1]{B. Sanderse}
\ead{B.Sanderse@cwi.nl}

\cortext[cor1]{Corresponding author}

\address[1]{Centrum Wiskunde \& Informatica,
            Science Park 123, 
            Amsterdam,
            The Netherlands}
\address[2]{Delft University of Technology, 
            Process and Energy, 
            Leeghwaterstraat 39, 
            Delft,
            The Netherlands}

\begin{abstract}
A novel hyper-reduction method is proposed that conserves kinetic energy and momentum for reduced order models of the incompressible Navier-Stokes equations. The main advantage of conservation of kinetic energy is that it endows the hyper-reduced order model (hROM) with a nonlinear stability property. The new method poses the discrete empirical interpolation method (DEIM) as a minimization problem and subsequently imposes constraints to conserve kinetic energy. Two methods are proposed to improve the robustness of the new method against error accumulation: oversampling and Mahalanobis regularization. Mahalanobis regularization has the benefit of not requiring additional measurement points. Furthermore, a novel method is proposed to perform structure-preserving temporal localization with the principle interval decomposition: new interface conditions are derived such that energy and momentum are conserved for a full time-integration instead of only during separate intervals. The performance of the new structure-preserving hyper-reduction methods and the structure-preserving temporal localization method is analysed using two convection-dominated test cases; a shear-layer roll-up and two-dimensional homogeneous isotropic turbulence. It is found that both Mahalanobis regularization and oversampling allow hyper-reduction of these test cases. Moreover, the Mahalanobis regularization provides comparable robustness while being more efficient than oversampling.
\end{abstract}


\end{frontmatter}

\section{Introduction}
\label{sec:introduction}

Computational fluid dynamics (CFD) has become an integral part of many modern engineering applications. The increase in computational power in recent decades has allowed engineers to model increasingly larger fluid dynamical systems. However, many modern applications are of a multi-query or real-time nature e.g.\ design optimization \cite{rozzabasic} and uncertainty quantification \cite{deliaquantification} or real-time control \cite{rowleymodel, noackreduced} and digital twin technology \cite{hartmannmodel}. These applications still pose prohibitively large computational costs. Reduced order models (ROMs) have been proposed as a solution to this problem. A ROM is a type of surrogate model that approximates the high-dimensional full scale model, or full order model (FOM), in a low-dimensional way by finding approximate formulations of involved quantities or operators. The low-dimensionality of the model consequently makes the ROM significantly cheaper to evaluate than the FOM.

Traditionally in the CFD community, these ROMs have been constructed by projecting the fluid dynamics equations of interest onto low-dimensional linear spaces obtained from the proper orthogonal decomposition (POD) algorithm \cite{turbulenceholmes, sirovichturbulence}, using either Galerkin \cite{sandersenonlinearly, rowleymodel, tezaurstablenon, Arunajatesanstable, Baronestable} or Petrov-Galerkin \cite{carlbergefficient, grimbergmesh, carlberggalerkin} projection. More recently, alternatives have also been explored. ROMs have been constructed without availability of a FOM by inferring the ROM from data using operator inference methods \cite{peherstorferdata, sharmahamiltonian, benneroperator, mcquarriedata}. Machine learning methods like convolutional autoencoders have been used in a projective sense \cite{leemodel, romornonlinear} and also inference methods \cite{muckeautoenc, maulikreduced} have been applied to obtain nonlinear low-dimensional approximations. Other nonlinear dimensionality reduction methods like quadratic manifolds \cite{barnettquadratic, geelenoperator} and diffusion maps \cite{sondaymanifold} have also been leveraged. However, the traditional POD-Galerkin methods remain powerful and this article will primarily deal with these methods and their natural extensions like the principal interval decomposition (PID) \cite{intervalboggaard, signalijzerman}.

Nevertheless, the traditional methods possess several limitations. For turbulent (and convection-dominated) flows that are of engineering interest it is well-known that linear, projection-based ROMs suffer from stability and accuracy issues \cite{sandersenonlinearly, fickstabilized, balajewiczstable, amsallemstable, Baronestable}. Efforts have been made to solve this issue, an overview is provided by \cite{fickstabilized, sandersenonlinearly}. A promising solution is structure-preservation, this entails constructing ROMs such that the underlying physics of fluid flows are respected. Especially conservation of kinetic energy is an important physical principle to uphold in a ROM with regards to stability as it bounds the norm of the solution \cite{sandersenonlinearly, afkhamconservative, entropychan, treuillemodel}. 

Current attempts at constructing structure-preserving ROMs of fluid flow can be classified into four categories. The first category is constrained optimization projection \cite{carlbergconservative, Scheinpreserving, Grimbergstability, bloniganmodel, huangmodel}; here model reduction is cast into an optimization problem and constrained to preserve structure. A second category is formed by symmetry-preservation \cite{sandersenonlinearly, afkhamconservative, entropychan, treuillemodel} where symmetries of the continuous analogues of ROM operators have been preserved resulting in the conservation of energy or entropy and associated stability. In \cite{gongstructure, chaturantabutstructure, polyugaeffort, hesthavenstructure, afkhamstructure} Hamiltonian physics are preserved in a low-dimensional setting, forming a third category. Finally, in \cite{mohebujjamanphysically, xiedata, loiseauconstrained, mohebujjamanenergy, balajewiczlow, akkaritime} physics-informed data-driven approaches using inference and machine learning methods have been adopted establishing a fourth category.

The idea of structure preservation is elegant but does not resolve directly the well-known issue that reducing nonlinear models using POD requires intermediate lifting of the reduced representation to the high-dimensional (FOM) spaces. Thus, although the ROM is low-dimensional, the computational effort to evaluate it is still high-dimensional, defeating its purpose. Methods to overcome this problem are referred to as hyper-reduction methods \cite{bruntondata}. For sufficiently simple equations exact hyper-reduction methods exist that eliminate computational dependence on the FOM dimensions \cite{sandersenonlinearly, ahmedbreaking}. This can be done when the underlying nonlinearity is of polynomial nature. Yet, these methods can become prohibitively expensive for larger ROMs as noted in \cite{sandersenonlinearly}. Hence, approximate hyper-reduction methods may be considered since these generally have better scaling properties. Examples of such approximate hyper-reduction methods are the discrete empirical interpolation method (DEIM) \cite{chaturantabutnonlinear, chaturantabutthesis, chaturantabutstate, barraulteim}, Gauss-Newton with approximated tensors (GNAT) \cite{carlberggnat, carlbergefficient} and energy-conserving sampling and weighting (ECSW) \cite{grimbergmesh, FarhatStructurepreserving, farhatdimensional}. 

Many existing hyper-reduction methods do not preserve the structure of the operators to which they are applied and thus stability can be lost. The field of model reduction of Hamiltonian systems has offered some solutions for this; \cite{chaturantabutstructure} proposes a structure-preserving DEIM variant for systems with Hamiltonian functionals with non-quadratic terms; \cite{wangstructure} improves this DEIM variant and \cite{miyatakestructure} proposes a structure-preserving DEIM variant for nonlinear Hamiltonian operators that preserves skew-symmetry. However, the incompressible Navier-Stokes equations considered in this work have a quadratic Hamiltonian functional in the inviscid case \cite{olverhamil, arnoldhamil} (kinetic energy) making the first two methods (\cite{chaturantabutstructure, wangstructure}) inapplicable. Furthermore, the approximate method proposed in \cite{miyatakestructure} scales computationally the same as the prohibitively expensive exact method used in \cite{sandersenonlinearly}. 

In this article we propose a novel structure-preserving DEIM that is capable to robustly, accurately and efficiently deal with convection-dominated flows described by the incompressible Navier-Stokes equations. The main idea will be to relax the DEIM interpolation requirement and use the resulting available degree of freedom to enforce energy conservation. Furthermore, we will propose two methods to further enhance the robustness of the structure-preserving DEIM. Namely, oversampling \cite{peherstorferstability, saibabarandomized} and the new approach of regularization using the Mahalanobis distance (which, to our knowledge, has not yet been applied to the DEIM). Finally, we will extend the feasibility of the hyper-reduced order model (hROM) with a newly proposed structure-preserving localization method based on the PID.

This article will be organized as follows. First, the governing equations will be introduced in \autoref{sec:prelim}, together with a structure-preserving FOM and a structure-preserving ROM. Following this, the structure-preserving hyper-reduction method will be proposed in \autoref{sec:hROM}. In \autoref{sec:time} the new structure-preserving interval decomposition approach for convection-dominated flows will be discussed. Finally, in \autoref{sec:results} the performance of the proposed structure-preserving hyper-reduction method and structure-preserving temporal localization method will be analysed using two convection-dominated test cases.

\section{Preliminaries: Navier-Stokes equations, FOM and ROM, energy equations} \label{sec:prelim}
\subsection{Incompressible Navier-Stokes equations and kinetic energy conservation}
In this work we consider the incompressible Navier-Stokes equations:
\begin{gather}
    \frac{\partial \boldsymbol{u}}{\partial t} + \nabla \cdot \left(\boldsymbol{u} \otimes \boldsymbol{u}\right)  =  -\nabla p + \nu \Delta \boldsymbol{u} + \boldsymbol{f}^b, \label{eq:ns}\\
    \nabla \cdot \boldsymbol{u} = 0, \label{eq:incomp}
\end{gather}
on a domain $\Omega \subset \mathbb{R}^d$ with periodic boundary conditions. Several quantities related to the flow will also be considered, namely kinetic energy $K(t) : \mathbb{R}^+ \rightarrow \mathbb{R}^+$, defined as:
\begin{equation}
    K(t) = \frac{1}{2} \left| \left| \boldsymbol{u} \right| \right|_{L^2}^2,
    \label{eq:totkin}
\end{equation}
and momentum $\boldsymbol{P}(t) : \mathbb{R}^+ \rightarrow \mathbb{R}^d$, defined as:
\begin{equation}
    \boldsymbol{P}(t) = \int_{\Omega} \boldsymbol{u} \hbox{ } d\Omega.
    \label{eq:totmom}
\end{equation}

Denoting $\mathcal{C}(\boldsymbol{u},\boldsymbol{u}) := \nabla \cdot \left(\boldsymbol{u} \otimes \boldsymbol{u}\right)$, $\mathcal{D}\boldsymbol{u} := \Delta \boldsymbol{u}$, $\mathcal{G}p := \nabla p$ and $\mathcal{M}\boldsymbol{u} := \nabla \cdot \boldsymbol{u}$, an evolution equation of the total kinetic energy can be found by differentiating \eqref{eq:totkin} and substituting \eqref{eq:ns}:
\begin{align*}
    \frac{dK}{dt} &= -\left< \boldsymbol{u}, \mathcal{C}(\boldsymbol{u},\boldsymbol{u})\right>_{L^2} -\left<\boldsymbol{u},\mathcal{G}p\right>_{L^2} + \nu \left<\boldsymbol{u},\mathcal{D}\boldsymbol{u} \right>_{L^2}.
\end{align*}
This evolution equation can be further simplified when the following is considered. Since the convection operator is skew-adjoint given condition \eqref{eq:incomp}, it holds that $\left< \boldsymbol{u}, \mathcal{C}(\boldsymbol{u},\boldsymbol{u})\right>_{L^2} = -\left<\mathcal{C}(\boldsymbol{u},\boldsymbol{u}), \boldsymbol{u}\right>_{L^2} = -\left< \boldsymbol{u}, \mathcal{C}(\boldsymbol{u},\boldsymbol{u})\right>_{L^2}$, meaning that $\left< \boldsymbol{u}, \mathcal{C}(\boldsymbol{u},\boldsymbol{u})\right>_{L^2} = 0$. Additionally, since the gradient operator and the divergence operator are each other's negated adjoints it can be stated that $\left<\boldsymbol{u},\mathcal{G}p\right>_{L^2} = - \left<\mathcal{M}\boldsymbol{u},p\right>_{L^2} = 0$ due to equation \eqref{eq:incomp}. Finally, the negative-definiteness of the diffusion operator can be employed to simplify the evolution equation of the total kinetic energy further to:
\begin{equation}
    \frac{dK}{dt} = -\nu \left| \left| \nabla \boldsymbol{u} \right| \right|_{L^2}^2 \leq 0.
    \label{eq:kin}
\end{equation}
Equation \eqref{eq:kin} implies that the kinetic energy, or the norm of the velocity field $\boldsymbol{u}$, is a monotonically decreasing quantity for periodic or homogeneous Dirichlet boundary conditions and is conserved in the inviscid case ($\nu = 0$). It will become clear soon that mimicking this property discretely will be crucial to nonlinear stability of the FOM and ROMs.

Furthermore, it follows from integrating equation \eqref{eq:ns} over the domain $\Omega$ that:
\begin{equation}
    \frac{d\boldsymbol{P}}{dt} = 0,
    \label{eq:mom}
\end{equation}
for periodic boundary conditions and with $\boldsymbol{f}^b = 0$. It is now our goal to \textit{mimic these conservation properties at the discrete and especially the reduced level}.

\subsection{Structure-preserving FOM}
Equation \eqref{eq:ns} is discretized with a finite volume method (FVM) on a staggered grid, resulting in a system of coupled ordinary differential equations (ODEs) complemented by a set of linear constraints:
\begin{gather}
    \Omega_h\frac{d \boldsymbol{u}_h}{dt} + C_h(\boldsymbol{u}_h) = -G_h \boldsymbol{p}_h + \nu D_h \boldsymbol{u}_h, \label{eq:hns}\\
    M_h \boldsymbol{u}_h = 0. \label{eq:hincomp}
\end{gather}
Here $\boldsymbol{u}_h(t) : \mathbb{R}^+ \rightarrow \mathbb{R}^N$ are the numerical velocity values arranged in a vector, $\Omega_h \in \mathbb{R}^{N \times N}$ is a diagonal matrix containing finite volume sizes, $C_h(\boldsymbol{u}_h) : \mathbb{R}^N \rightarrow \mathbb{R}^N$ is the spatial discretization of the nonlinear convection operator, $G_h \in \mathbb{R}^{N \times N_p}$ is the spatial discretization of the gradient operator, $\boldsymbol{p}_h(t) : \mathbb{R}^+ \rightarrow \mathbb{R}^{N_p}$ are the numerical pressure values arranged in a vector, $D_h \in \mathbb{R}^{N \times N}$ is the spatial discretization of the diffusion operator, $M_h \in \mathbb{R}^{N_p \times N}$ is the spatial discretization of the divergence operator, $N = N_u + N_v$ is the total number of velocity unknowns and $N_u$ and $N_v$ are the numbers of velocity unknowns in the $x$ and $y$ directions respectively (for the case of two-dimensional domains) and $N_p$ is the number of pressure unknowns. Relations \eqref{eq:hns}-\eqref{eq:hincomp} are in turn complemented by a vector of initial conditions $\boldsymbol{u}_0 = \boldsymbol{u}_h(0)$ and suitable boundary conditions. The choice of boundary conditions for the flows considered in this work are periodic.

We use the discretization as described in \cite{sandersdissertation, sandersenonlinearly} such that the following three important properties of the continuous operators are inherited by their spatial discretizations. First, the convection operator, which can be written in quasi-linear form as $C_h(\boldsymbol{u}_h) = \widetilde{C}_h(\boldsymbol{u}_h) \boldsymbol{u}_h$, is skew-symmetric:
\begin{equation}
    \widetilde{C}_h(\boldsymbol{u}_h) = - \widetilde{C}_h(\boldsymbol{u}_h)^T.
    \label{eq:chssymm}
\end{equation}
Second, the duality between gradient and divergence operators holds:
\begin{equation}
    G_h = -M_h^T.
    \label{eq:hdivgradrel}
\end{equation}
Third, the diffusive operator is symmetric negative-definite, allowing the decomposition:
\begin{equation}
     D_h = - Q_h^T Q_h.
     \label{eq:dhsymm}
\end{equation}
We remark that the conservation properties discussed in this article are not specific to the method described in \cite{sandersdissertation, sandersenonlinearly}, any other finite volume discretization of \eqref{eq:ns}-\eqref{eq:incomp} satisfying the above properties can also be used.

Using these properties and the telescoping property of the FVM, discretization \eqref{eq:hns}-\eqref{eq:hincomp} conserves discrete analogues of mass, total momentum and total kinetic energy. To show this, we first define discrete total kinetic energy and momentum on the basis of an inner product. To this end, the $\Omega_h$-inner product is introduced. The $\Omega_h$-inner product and its induced norm are defined by:
\begin{equation*}
    \left<\boldsymbol{u},\boldsymbol{v}\right>_{\Omega_h} := \left<\boldsymbol{u}, \Omega_h \boldsymbol{v}\right>, \quad \left|\left| \boldsymbol{u} \right|\right|_{\Omega_h} := \sqrt{\left<\boldsymbol{u},\boldsymbol{u}\right>_{\Omega_h}},
\end{equation*}
where $\left<\cdot,\cdot\right>$ denotes the standard Euclidean inner product. After deriving an evolution equation for the discrete total kinetic energy it will be clear why this is a natural choice. Using the $\Omega_h$-inner product, the discrete total kinetic energy $K_h(t) : \mathbb{R}^+ \rightarrow \mathbb{R}^+$ is defined as:
\begin{equation*}
    K_h(t) := \frac{1}{2} \left| \left| \boldsymbol{u}_h \right| \right|_{\Omega_h}^2.
\end{equation*}
The discrete total momentum $\boldsymbol{P}_h(t) : \mathbb{R}^+ \rightarrow \mathbb{R}^d$ is defined as:
\begin{equation*}
    \left(\boldsymbol{P}_h(t)\right)_i = \left<\boldsymbol{e}_i,\boldsymbol{u}_h\right>_{\Omega_h},
\end{equation*}
where $\boldsymbol{e}_i \in \mathbb{R}^N$ is a vector of ones at the vector indices where $\boldsymbol{u}_h$ has velocity components in direction $i \in \{1,...,d\}$ and zeros elsewhere.

An evolution equation for discrete kinetic energy is found by temporal differentiation of $K_h(t)$:
\begin{align}
    \frac{dK_h}{dt} = \frac{1}{2} \frac{d}{dt} \left<\boldsymbol{u}_h,\boldsymbol{u}_h\right>_{\Omega_h} 
    &= -\left<\boldsymbol{u}_h, \widetilde{C}_h(\boldsymbol{u}_h) \boldsymbol{u}_h \right> - \left<\boldsymbol{u}_h, G_h \boldsymbol{p}_h \right> + \nu \left<\boldsymbol{u}_h, D_h \boldsymbol{u}_h\right>\nonumber \\
    &= - \nu \left| \left|Q_h \boldsymbol{u}_h \right| \right|^2 \leq 0, \label{eq:kinh}
\end{align}
where $||\cdot||$ denotes the Euclidean norm. The simplification in the second line follows from the application of properties \eqref{eq:chssymm}-\eqref{eq:dhsymm}. Indeed, evolution equation \eqref{eq:kinh} is a discrete analogue to equation \eqref{eq:kin} and shows that $K_h$, and so the norm of $\boldsymbol{u}_h$, is a monotonically decreasing quantity. The solution of the semi-discrete system of ODEs \eqref{eq:hns} can thus be bounded from above by:
\begin{equation*}
    \left| \left| \boldsymbol{u}(t) \right|\right|_{\Omega_h} \leq \left| \left| \boldsymbol{u}_0 \right|\right|_{\Omega_h},
\end{equation*}
\textit{guaranteeing stability} of $\boldsymbol{u}_h(t)$.

It is also a well-known property of the FVM, referred to as the telescoping property \cite{wesselingprinciples}, that:
\begin{align}
    \left(\frac{d\boldsymbol{P}_h}{dt}\right)_i = \frac{d}{dt}\left<\boldsymbol{e}_i,\boldsymbol{u}_h\right>_{\Omega_h} &= -\left<\boldsymbol{e}_i, \widetilde{C}_h(\boldsymbol{u}_h) \boldsymbol{u}_h \right> - \left<\boldsymbol{e}_i, G_h \boldsymbol{p}_h \right> + \nu \left<\boldsymbol{e}_i, D_h \boldsymbol{u}_h\right> \nonumber \\
    &= 0, \quad \forall i \in \{1,...,d\}. \label{eq:momh}
\end{align}
This constitutes a discrete analogue to \eqref{eq:mom} and shows that discrete total momentum is also conserved.

\subsection{Structure-preserving ROM}
To construct a reduced order model as in \cite{sandersenonlinearly} we make the assumption that for any $t \in [0,T]$ there are elements $\boldsymbol{u}_r(t)$ of a low-dimensional linear subspace $\mathcal{V} \subset \mathbb{R}^N$ that accurately approximate $\boldsymbol{u}_h(t)$. Here, we denote $r := \text{dim}(\mathcal{V})$ and the low dimensionality of $\mathcal{V}$ implies that $r \ll N$. Furthermore, we consider $\mathcal{V} \subset \text{ker}(M_h)$ such that all $\boldsymbol{u}_r(t)$ automatically satisfy condition \eqref{eq:hincomp}. Let $\Phi \in \mathbb{R}^{N \times r}$ be a POD basis for $\mathcal{V}$ that is orthonormal in the $\Omega_h$-inner product, i.e. $\left<\Phi_{,i},\Phi_{,j}\right>_{\Omega_h} = \delta_{ij}$ \footnote{We use commas in the notation $(A_{,j})$ to the denote the $j^{\text{th}}$ column of matrix $A$, similarly $(A_{i,})$ denotes the $i^{\text{th}}$ row.}, then we write:
\begin{equation}
    \boldsymbol{u}_h(t) \approx \boldsymbol{u}_r(t) := \Phi \boldsymbol{a}(t) \in \mathcal{V}.
    \label{eq:romappr}
\end{equation}
Here $\boldsymbol{a}(t) : \mathbb{R}^+ \rightarrow \mathbb{R}^r$ are the generalized coordinates in $\mathcal{V}$ associated to the POD basis $\Phi$. Substituting this approximation into the FOM \eqref{eq:hincomp} we obtain an overdetermined system.
 To obtain a solvable system, we test the overdetermined system against the POD modes in $\Phi$ and impose the Galerkin condition. This results in a pressure-free\footnote{Note that since $\mathcal{V} \subset \text{ker}(M_h)$, we can write $\Phi^T G_h p_h = -(M_h^T \Phi)^T p_h = 0$, making the ROM velocity-only.} ROM:
\begin{equation}
     \frac{d\boldsymbol{a}}{dt} = - \widetilde{C}_r( \boldsymbol{a})\boldsymbol{a} + \nu D_r \boldsymbol{a},
    \label{eq:rom}
\end{equation}
where $\widetilde{C}_r(\boldsymbol{a}) := \Phi^T\widetilde{C}_h(\Phi \boldsymbol{a})\Phi$ and $D_r := \Phi^TD_h\Phi$.

In order to consider structure-preservation we first define the following quantities. A reduced order representation of kinetic energy is defined as: 
\begin{equation}
    K_r(t) := \frac{1}{2}||\boldsymbol{u}_r||_{\Omega_h}^2 = \frac{1}{2}||\boldsymbol{a}||^2.
    \label{eq:Kr}
\end{equation}
For total momentum a reduced order representation is:
\begin{equation*}
    \left(\boldsymbol{P}_r(t)\right)_i := \left<\boldsymbol{e}_i, \boldsymbol{u}_r\right>_{\Omega_h} = \left<\Phi^T\Omega_h\boldsymbol{e}_i,\boldsymbol{a}\right>. 
\end{equation*}

 Noting that symmetry properties \eqref{eq:chssymm} and \eqref{eq:dhsymm} are still satisfied by the reduced operators in \eqref{eq:rom}, the evolution equation for $K_r(t)$ follows by differentiating \eqref{eq:Kr}:
\begin{align}
    \frac{dK_r}{dt} = \frac{1}{2} \frac{d}{dt} \left<\boldsymbol{a},\boldsymbol{a}\right>  
    &= - \left<\boldsymbol{a}, \widetilde{C}_r( \boldsymbol{a})\boldsymbol{a}\right> + \nu \left<\boldsymbol{a}, D_r \boldsymbol{a}\right> \nonumber \\
    &= - \nu \left| \left| Q_r \boldsymbol{a} \right| \right|^2 \leq 0,
    \label{eq:kinrevo2}
\end{align}
where $Q_r := Q_h \Phi$. This equation is the ROM analogue of equation \eqref{eq:kin} and shows that the norm of the generalized coordinates satisfies:
\begin{equation*}
    \left|\left|\boldsymbol{a}(t)\right|\right| \leq \left|\left|\boldsymbol{a}_0\right|\right|,
\end{equation*}
with equality for $\nu=0$. 

To conserve total momentum at the reduced level we require $\boldsymbol{e}_i \in \mathcal{V} \hbox{ } \forall i \in \{1,...,d\}$. This is done by explicitly selecting $\boldsymbol{e}_i /||\boldsymbol{e}_i||_{\Omega_h}$ as the first $d$ POD modes in $\Phi$. An evolution equation for $\boldsymbol{P}_r(t)$ is then found as:
\begin{align} 
    \left(\frac{d\boldsymbol{P}_r}{dt}\right)_i = \frac{d}{dt}\left<\Phi^T\Omega_h\boldsymbol{e}_i,\boldsymbol{a}\right> &= -\left<\Phi^T\Omega_h\boldsymbol{e}_i, \widetilde{C}_r(\boldsymbol{a}) \boldsymbol{a}\right> + \nu \left<\Phi^T\Omega_h\boldsymbol{e}_i, D_r \boldsymbol{a} \right> \nonumber \\ 
    &= -\left<\boldsymbol{e}_i, \widetilde{C}_h(\Phi \boldsymbol{a})\Phi \boldsymbol{a}\right> + \nu \left<\boldsymbol{e}_i, D_h \Phi \boldsymbol{a} \right> \\
    &= 0. \nonumber
\end{align}
In the second line the telescoping property of the FOM operators was evoked.

Finally, a POD basis $\Phi$ satisfying all required criteria ($\text{span}(\Phi) \subset \text{ker}(M_h)$, $\Omega_h$-orthonormality and containing all $\boldsymbol{e}_i /||\boldsymbol{e}_i||_{\Omega_h}$ in its columns) is constructed from snapshot data using an altered POD method as described in \cite{sandersenonlinearly}. The FOM and the ROM can both be integrated in time with energy-conserving Runge-Kutta methods to conserve energy fully discretely (see \cite{sandersenonlinearly, sanderseenergy} for further details). 

\section{A structure-preserving discrete empirical interpolation method} \label{sec:hROM}
\subsection{Introduction to DEIM}
The ROM \eqref{eq:rom} as constructed in the previous section is $r$-dimensional, but evaluating the convective terms in a naive fashion still requires a computational effort that scales with the FOM dimension $N$. One approach to alleviate this issue is the exact tensor decomposition. This method results in an exact low-dimensional representation of the convection operator but scales with $\mathcal{O}(r^3)$, which becomes computationally expensive for realistic values of $r$ \cite{sandersenonlinearly}. A more efficient alternative approach, which will be considered here, is the discrete empirical interpolation method (DEIM) \cite{chaturantabutnonlinear}. Using the DEIM we will construct a so-called hyper-reduced order model (hROM).

The DEIM approximates $C_h(\boldsymbol{u})$ with elements from a low-dimensional linear subspace $\mathcal{M}_d \subset \mathbb{R}^N$. Here, we denote $m := \text{dim}(\mathcal{M}_d)$ and the low dimensionality of $\mathcal{M}_d$ implies that $m \ll N$. Let $M \in \mathbb{R}^{N \times m}$ be an orthonormal basis for $\mathcal{M}_d$ constructed from snapshot data of the operator using the POD algorithm, where the snapshot data is gathered in a matrix $\Xi \in \mathbb{R}^{N\times n_s}$ given as:
\begin{equation}
    \Xi = \left[ C_h(\boldsymbol{u}_h(t^0)), C_h(\boldsymbol{u}_h(t^0 + \Delta t)), ... , C_h(\boldsymbol{u}_h(t^0 + (n_s-1) \Delta t)) \right].
\end{equation}
The basis $M$ thus satisfies $\left<M_{,i},M_{,j}\right> = \delta_{ij}$. Note the distinction between the basis $M$ and the discrete divergence operator $M_h$. The DEIM approximation then takes the form:
\begin{equation}
    C_h(\Phi\boldsymbol{a}) \approx M \boldsymbol{c}(\boldsymbol{a}) \in \mathcal{M}_d,
    \label{eq:deimappr}
\end{equation}
where $\boldsymbol{c}(\boldsymbol{a}) : \mathbb{R}^r \rightarrow \mathbb{R}^m$ are the generalized coordinates in $\mathcal{M}_d$ associated to basis $M$ and are referred to as DEIM coordinates. Using the DEIM approximation the ROM takes the form:
\begin{equation}
    \frac{d\boldsymbol{a}}{dt} = - \Phi^T M\boldsymbol{c}(\boldsymbol{a}) + \nu D_r\boldsymbol{a}.
    \label{eq:deimrom}
\end{equation}
The linear nature of the approximation allows the Galerkin projection $\Phi^TM$ to be precomputed in an offline step. Hence, with the values of the DEIM coordinates available, the ROM can be evaluated in a completely low-dimensional fashion. 

However, the problem \eqref{eq:deimappr} to calculate the DEIM coordinates is overdetermined. Furthermore, the least-squares solution to system \eqref{eq:deimappr}, $\boldsymbol{c} = M^T C_h(\Phi \boldsymbol{a})$, requires an expensive projection operation. To solve this, the DEIM bases its approximation on only an $m$-dimensional subset of evaluations of $C_h(\Phi\boldsymbol{a})$ at points on the grid, referred to as the measurement points, and denoted by $\mathcal{P}$. More specifically, a measurement matrix $P \in \{0,1\}^{N\times m}$ is defined that consists of selected columns of the $N\times N$ identity matrix corresponding to the measurement points. To calculate the DEIM coordinates the following system is then solved:
\begin{equation}
    P^TC_h(\Phi\boldsymbol{a}) = P^TM\boldsymbol{c}.
    \label{eq:corrdeim}
\end{equation}
The formal solution to this system is $\boldsymbol{c} = (P^TM)^{-1}P^TC_h(\Phi\boldsymbol{a})$. A greedy measurement point selection procedure that ensures that $(P^TM)^{-1}$ exists and that provides near-optimal approximations of $C_h(\Phi \boldsymbol{a})$ was proposed in \cite{chaturantabutnonlinear}. Using this method, the DEIM coordinates can be determined using only $m$ evaluations of $C_h(\Phi \boldsymbol{a})$ and $(P^TM)^{-1}$ can be precomputed. This procedure results in a completely low-dimensional ROM.

We will analyse the conservation properties of this `conventional' DEIM approach. Because the telescoping property of operators discretized using the FVM is a linear property, it is conserved under application of the POD algorithm. This may be shown by analysing the eigenvectors of the correlation matrix $\Xi \Xi^T$. Using this property it can be shown that momentum is conserved for a hROM using the DEIM:
\begin{align} 
    \left(\frac{d\boldsymbol{P}_r}{dt}\right)_i = \frac{d}{dt}\left<\Phi^T\Omega_h\boldsymbol{e}_i,\boldsymbol{a}\right> &= -\left<\Phi^T\Omega_h\boldsymbol{e}_i, \Phi^TM\boldsymbol{c}(\boldsymbol{a})\right> + \nu \left<\Phi^T\Omega_h\boldsymbol{e}_i, D_r \boldsymbol{a} \right> \nonumber \\ 
    &= -\left<\boldsymbol{e}_i, M\boldsymbol{c}(\boldsymbol{a})\right> + \nu \left<\boldsymbol{e}_i, D_h \Phi \boldsymbol{a} \right> \label{eq:momconsrom} \\
    &= 0. \nonumber
\end{align}

We now arrive at the key issue addressed in this article. This issue is that \textit{the hROM \eqref{eq:deimrom} does not conserve reduced total kinetic energy}, i.e.\ it can occur that:
\begin{align}
    \frac{dK_r}{dt} = \frac{1}{2} \frac{d}{dt} \left<\boldsymbol{a},\boldsymbol{a}\right>  
    &= - \left<\boldsymbol{a}, \Phi^T M\boldsymbol{c}(\boldsymbol{a})\right> - \nu \left| \left| Q_r \boldsymbol{a} \right| \right|^2 > 0,
    \label{eq:kinrevo3}
\end{align}
because generally $\left<\boldsymbol{a}, \Phi^T M\boldsymbol{c}(\boldsymbol{a})\right>$ does not equal zero. This is because the DEIM-interpolated convection operator:
\begin{equation}
    \Phi^T M (P^TM)^{-1} P^T \widetilde{C}_h(\Phi \boldsymbol{a})\Phi,
    \label{eq:quasiop}
\end{equation}
is no longer skew-symmetric. This can lead to instability as the norm of the generalized coordinates $\boldsymbol{a}$ is not bounded from above. An approach to retain the skew-symmetry of a DEIM approximation and therefore the energy-conservation property of the ROM has been proposed in \cite{miyatakestructure}. However, this method scales equivalently to the exact tensor decomposition and is therefore considered infeasible for realistic fluid dynamics applications. We will present a new approach to make the hROM energy-conserving, while also retaining the momentum conservation property.

\subsection{A novel structure-preserving DEIM}
The method we propose will be named the constrained least-squares discrete empirical interpolation method (CLSDEIM). It is based on the realization that skew-symmetry of the interpolated quasi-linear convection operator \eqref{eq:quasiop} is a sufficient \textit{but not a necessary condition} for energy-conservation. Instead, a necessary condition to conserve reduced total kinetic energy is:
\begin{equation}
    \left<\boldsymbol{a}, \Phi^T M \boldsymbol{c}(\boldsymbol{a})\right> = 0.
    \label{eq:deimcond}
\end{equation}
Condition \eqref{eq:deimcond} can be satisfied even if the Galerkin-projected DEIM operator $\Phi^TM\boldsymbol{c}(\boldsymbol{a}) : \mathbb{R}^r \rightarrow \mathbb{R}^r$ is not skew-symmetric and, when satisfied, results in the correct reduced kinetic energy evolution equation \eqref{eq:kinrevo2}. The new idea of the CLSDEIM is to enforce this condition by posing the DEIM as a constrained minimization problem.

The conventional DEIM finds the DEIM coordinates $\boldsymbol{c}$ by minimizing the Euclidean norm between the nonlinearity and the DEIM approximation in the measurement space. The Euclidean norm can be considered as minimized since the difference in $\mathcal{P}$ i.e. $P^T C_h(\Phi \boldsymbol{a}) - P^T M \boldsymbol{c}$, is zero. The idea of the proposed CLSDEIM is to employ this view of DEIM as a minimization problem and constrain it to take place over the set $\mathcal{F}(\boldsymbol{a})$ of DEIM approximations satisfying condition \eqref{eq:deimcond}, defined using the DEIM coordinates as:
\begin{equation*}
    \mathcal{F}(\boldsymbol{a}) := \{ \boldsymbol{c} \in \mathbb{R}^m \hbox{ } | \hbox{ } \boldsymbol{a}^T \Phi^T M \boldsymbol{c} = 0 \}.
\end{equation*}
The set $\mathcal{F}(\boldsymbol{a})$ is referred to as the feasible set. As DEIM approximations with $\boldsymbol{c}(\boldsymbol{a}) \in \mathcal{F}(\boldsymbol{a})$ satisfy condition \eqref{eq:deimcond}, the CLSDEIM produces approximations that conserve reduced total kinetic energy. The constrained minimization problem to find the DEIM coordinates $\boldsymbol{c}$ will be posed as the following linearly constrained least-squares problem:
\begin{equation}
    \boxed{\boldsymbol{c}(\boldsymbol{a}) = \argmin_{\widetilde{\boldsymbol{c}} \in \mathbb{R}^m} \left|\left|P^T C_h(\Phi \boldsymbol{a}) - P^T M \widetilde{\boldsymbol{c}} \right|\right|^2 \quad \text{s.t.} \quad \boldsymbol{a}^T \Phi^T M \widetilde{\boldsymbol{c}} = 0.}
    \label{eq:lsdeim}
\end{equation}
This means the CLSDEIM relaxes condition \eqref{eq:corrdeim} of exact correspondence in $\mathcal{P}$ between the FOM's convection operator and the DEIM approximation imposed by the conventional DEIM. Rather, the CLSDEIM minimizes the difference between the approximation and the FOM operator in $\mathcal{P}$ while simultaneously constraining the approximation to be energy-conserving.

Considering the objective function and the geometric interpretation of the DEIM \cite{chaturantabutthesis, chaturantabutnonlinear}, the CLSDEIM can be interpreted geometrically as an oblique projection of $C_h(\Phi \boldsymbol{a})$ on the subspace $\mathcal{M}_\mathcal{F} \subset \mathbb{R}^N$ of all DEIM approximations $M\boldsymbol{c}$ with $\boldsymbol{c} \in \mathcal{F}(\boldsymbol{a})$. The subspace $\mathcal{M}_\mathcal{F}$ is defined as:
\begin{equation*}
    \mathcal{M}_\mathcal{F} = \mathcal{M}_d \cap \text{ker}(\left<\Phi\boldsymbol{a},\cdot\right>),
\end{equation*}
where $\left<\Phi\boldsymbol{a},\cdot\right> : \mathbb{R}^N \rightarrow \mathbb{R}$ is the functional taking the Euclidean inner product with $\Phi \boldsymbol{a}$. Particularly, $\mathcal{M}_\mathcal{F}$ are all vectors in $\mathbb{R}^N$ that can be written as a linear combination of the columns of $M$ and satisfy condition \eqref{eq:deimcond}. Since $\mathcal{M}_\mathcal{F}$ is the intersection between two linear subspaces of $\mathbb{R}^N$ it is also a linear subspace of $\mathbb{R}^N$. Contrary to the conventional DEIM, the CLSDEIM also projects obliquely through the measurement space $\mathcal{P}$. Hence, the orthogonal projection of the DEIM residual $\boldsymbol{r}(t) \in \mathbb{R}^N$ on $\mathcal{P}$, as given by:
\begin{equation*}
    P^T\boldsymbol{r}(t) = P^T \left[C_h(\Phi \boldsymbol{a}(t)) - M\boldsymbol{c}(t)\right],
\end{equation*}
is generally not zero. However, the CLSDEIM projector $\Pi_{\mathcal{M}_\mathcal{F}} : \mathbb{R}^N \rightarrow \mathcal{M}_\mathcal{F}$ does project $C_h(\Phi \boldsymbol{a})$ onto $\mathcal{M}_\mathcal{F}$ such that the orthogonally projected residual $P^T\boldsymbol{r}(t)$ is minimal in the Euclidean norm. This statement follows exactly from the formulation of the CLSDEIM minimization problem \eqref{eq:lsdeim}.

Furthermore, the CLSDEIM basis $M$ and the measurement space $\mathcal{P}$ can simply be found following the procedures of the conventional DEIM algorithm. Indeed, as we are using the conventional DEIM basis the reduced total momentum will also remain a conserved quantity for the CLSDEIM as was shown in equation \eqref{eq:momconsrom}. 

The constrained minimization problem \eqref{eq:lsdeim} is solved using the method of Lagrange multipliers \cite{boydconvex}. The Lagrangian $\mathcal{L}(\boldsymbol{c},\lambda) : \mathbb{R}^m \times \mathbb{R} \rightarrow \mathbb{R}$ is defined as:
\begin{equation*}
    \mathcal{L}(\boldsymbol{c},\lambda) = \left|\left|P^TC_h(\Phi \boldsymbol{a}) - P^T M \boldsymbol{c} \right|\right|^2 + \lambda \boldsymbol{a}^T\Phi^TM\boldsymbol{c},
\end{equation*}
where $\lambda \in \mathbb{R}$ is a Lagrange multiplier. Taking partial derivatives of the Lagrangian and setting them to zero leads to the following system for the optima:
\begin{equation}
    \begin{bmatrix}
                   2 (P^T M)^T P^T M &  (\boldsymbol{a}^T \Phi^T M)^T \\
                   \boldsymbol{a}^T \Phi^T M & 0 
    \end{bmatrix}
    \begin{bmatrix}
                   \boldsymbol{c}_o \\
                   \lambda_o
    \end{bmatrix} =
    \begin{bmatrix}
                   2(P^T M)^T P^T C_h(\Phi \boldsymbol{a}) \\
                   0
    \end{bmatrix}.
    \label{eq:lagrsys}
\end{equation}
Note that this matrix is symmetric and has constant coefficients with the exception of the last row and column which depend on $\boldsymbol{a}$. Solving \eqref{eq:lagrsys} can be done explicitly using the inverse of a symmetric $2\times 2$ block matrix  \cite{luinverses}:
\begin{equation*}
    \begin{bmatrix}
                   A & B \\
                   B^T & 0 
    \end{bmatrix}^{-1}
    =
    \begin{bmatrix}
                   A^{-1} - A^{-1} B (B^T A^{-1} B)^{-1} B^T A^{-1}   &  A^{-1} B(B^T A^{-1}B)^{-1} \\
                   (B^T A^{-1} B)^{-1} B^T A^{-1}                 &  -(B^T A^{-1} B)^{-1}
    \end{bmatrix},
\end{equation*}
where $A$ is symmetric. Taking:
\begin{align*}
    &A := 2(P^T M)^T P^T M &&\in \mathbb{R}^{m\times m}\\
    &B := (\boldsymbol{a}^T \Phi^T M)^T = M^T\Phi \boldsymbol{a} &&\in \mathbb{R}^{m \times 1},
\end{align*}
the DEIM coordinates $\boldsymbol{c}_o$ solving \eqref{eq:lsdeim} are determined by:
\begin{equation*}
    \boldsymbol{c}(\boldsymbol{a}) = \left(A^{-1} - A^{-1} B (B^T A^{-1} B)^{-1} B^T A^{-1}\right) 2(P^T M)^T P^T C_h(\Phi \boldsymbol{a}).
\end{equation*}
We use the notation $B := \boldsymbol{b}(\boldsymbol{a})$ to explicitly denote the dependence of $B$ on $\boldsymbol{a}$ and the fact that it can be treated as a vector. Since $B^T A^{-1} B = \boldsymbol{b}(\boldsymbol{a})^T A^{-1} \boldsymbol{b}(\boldsymbol{a})$ and $B^TA^{-1} 2(P^T M)^T P^T C_h(\Phi \boldsymbol{a}) =  \boldsymbol{b}(\boldsymbol{a})^T(P^TM)^{-1} P^T C_h(\Phi \boldsymbol{a})$ are simply scalars, the final expression for $\boldsymbol{c}_o$ is:
\begin{equation}
    \boxed{\text{CLSDEIM:} \qquad \boldsymbol{c}(\boldsymbol{a}) = (P^T M)^{-1} P^T C_h(\Phi \boldsymbol{a}) - \frac{\boldsymbol{b}(\boldsymbol{a})^T (P^T M)^{-1} P^T C_h(\Phi \boldsymbol{a})}{(\boldsymbol{b}(\boldsymbol{a})^T A^{-1} \boldsymbol{b}(\boldsymbol{a}))} A^{-1} \boldsymbol{b}(\boldsymbol{a}).}
    \label{eq:clsdeim}
\end{equation}
The DEIM coordinates found from equation \eqref{eq:clsdeim} result in a DEIM approximation $M\boldsymbol{c}$ satisfying condition \eqref{eq:deimcond} and consequently also equation \eqref{eq:kinrevo2}. All terms in \eqref{eq:clsdeim} are either inner products or matrix-vector products between low-dimensional vectors or matrices and vectors respectively, hence the CLSDEIM can be solved in a low-dimensional fashion. Note that the first term on the right hand side of equation \eqref{eq:clsdeim} is the solution to equation \eqref{eq:corrdeim}. Hence, equation \eqref{eq:clsdeim} can be interpreted as the conventional DEIM with a correction term. The correction term projects the conventional DEIM coordinates on $\text{ker}(\left<\boldsymbol{b}(\boldsymbol{a}),\cdot\right>)$ such that condition \eqref{eq:deimcond} is satisfied and minimization problem \eqref{eq:lsdeim} is solved.

\begin{rmk}\label{rmk:rk}
Both implicit and explicit Runge-Kutta (RK) methods will be considered to integrate the proposed hROMs. To achieve exact energy conservation (for inviscid flows), \textit{implicit} energy-conserving RK methods need to be used \cite{sanderseenergy,sandersenonlinearly}. Using such methods and structure-preserving hyper-reduction, the change in reduced total kinetic energy $K_r$ is given by:
\begin{equation}
    \boxed{K_r^{n+1} - K_r^n = -\Delta t \sum_{i=1}^s b_i \nu \left| \left| Q_r \boldsymbol{A}_i \right|\right|^2,}
    \label{eq:rkkinr2}
\end{equation}
which is the time-discrete equivalent of \eqref{eq:kinrevo2}. Here, $s \in \mathbb{N}$ and $b_i \in \mathbb{R}$ are parameters of the given RK method and $\boldsymbol{A}_i \in \mathbb{R}^r$ are the generalized coordinates at the $i^{\text{th}}$ RK stage. In the inviscid case it can be seen from \eqref{eq:rkkinr2} that the hROMs conserve reduced kinetic energy. In the viscous case the resulting hROMs will be nonlinearly stable when $b_i \geq 0 \hbox{ } \forall \hbox{ } i \in \{1,...,s\}$, which holds for the Gauss-Legendre RK methods (see \cite{sandersenonlinearly, sanderseenergy} for further details).
\end{rmk}

\subsection{Preventing overfitting errors}
Interpolation-based methods like the DEIM and CLSDEIM can be prone to overfitting when noise is present \cite{hastieelements}. It was shown in \cite{peherstorferstability} that adding Gaussian noise $\epsilon \in \mathbb{R}^m$ to the measurements, i.e.\ $P^TC_h(\boldsymbol{u}) + \epsilon$, results in error bounds on the DEIM approximation of $C_h(\boldsymbol{u})$ that grow as $\mathcal{O}(\sqrt{m})$. Similar overfitting errors may occur in the presence of numerical errors in $\boldsymbol{a}$ as we will show using experiments later. As a result of such overfitting errors, numerical errors in the solution may accumulate, further harming the capability of the DEIM to reconstruct the correct $C_h(\Phi \boldsymbol{a})$ of the following time steps. This is especially notable when the interpolation procedure is carried out many times, which is the case when solving time-dependent problems. We remark that this can happen even in the case of fully-discrete energy-conserving schemes since energy conservation is a restriction on the norm of the solution and not on its accuracy. In this case, overfitting errors typically manifest themselves in non-convergence of the nonlinear solver. To address overfitting we therefore propose two further stabilization approaches. The first one is oversampling, which has recently gained significant attention \cite{peherstorferstability, saibabarandomized}. The second approach is regularization, which is a well-known technique from optimization and has thus far, to the author's knowledge, not been applied to the DEIM. This latter aproach has the benefit that additional measurement points are not necessary and therefore reduces the computational burden compared to oversampling.

\subsubsection{Oversampling}
When using oversampling we replace the interpolation of DEIM \eqref{eq:corrdeim} and CLSDEIM \eqref{eq:lsdeim} with a regression procedure. This is achieved by considering a larger measurement space than the dimension of the DEIM space $\mathcal{M}_d$, thus $m_p := \text{dim}\left(\mathcal{P}\right) > m$. As a result, the measurement matrix becomes $P \in \{0,1\}^{N \times m_p}$. The formulation of the underlying optimization problem remains the same as optimization problem \eqref{eq:lsdeim}, but now the larger measurement matrix is used. Due to the fact that we are trying to fit a linear combination of $m$ DEIM modes to $m_p$ measurements with $m_p > m$, the resulting DEIM approximation $M\boldsymbol{c}$ is less sensitive to noise in measurements of $C_h(\Phi \boldsymbol{a})$ in $\mathcal{P}$. This was also analysed in \cite{peherstorferstability}, where it was shown that oversampling using randomized measurement points resulted in a contribution of noise to the error bound of the approximation of $C_h(\boldsymbol{u})$ that scales with $\mathcal{O}\left(\sqrt{m/m_p}\right)$. Hence, overfitting errors as a result of noise can be mitigated by choosing $m_p$ sufficiently large compared to $m$. We will refer to CLSDEIM being extended with oversampling as OCLSDEIM. The DEIM coordinates as calculated using the OCLSDEIM require solving optimization problem \eqref{eq:lsdeim} with the new measurement matrix $P$. The solution procedure of this problem is completely analogous to that of the CLSDEIM \eqref{eq:lsdeim} and the final expression for $\boldsymbol{c}_o$ is:
\begin{equation}
    \boxed{\text{OCLSDEIM:} \qquad \boldsymbol{c}(\boldsymbol{a}) = (P^T M)^{\dagger} P^T C_h(\Phi \boldsymbol{a}) - \frac{\boldsymbol{b}(\boldsymbol{a})^T (P^T M)^{\dagger} P^T C_h(\Phi \boldsymbol{a})}{(\boldsymbol{b}(\boldsymbol{a})^T A^{-1} \boldsymbol{b}(\boldsymbol{a}))} A^{-1} \boldsymbol{b}(\boldsymbol{a}),}
    \label{eq:oclsdeim}
\end{equation}
where instead of $(P^T M)^{-1}$ the Moore-Penrose pseudoinverse $(P^T M)^{\dagger}$ appears.

The measurement point selection procedure will be as follows. The first $m$ measurement points are found using the classic DEIM procedure described in \cite{chaturantabutnonlinear}. The next $m_p - m$ points are generated from a uniform random distribution where care is taken to not select repeating indices. This procedure strikes a good balance between efficiency and accuracy and is simple to implement. Alternatives are possible, see for example the following articles for deterministic approaches to measurement point selection: \cite{peherstorferstability, saibabarandomized, zimmermannaccelerated, manohardatadriven}.

\subsubsection{Generalized Tikhonov regularization} 
When overfitting occurs, DEIM modes interpolate measurements containing noise or numerical errors resulting from time integration, negatively effecting accuracy in subsequent time steps. Although the DEIM measurement points are chosen to minimize $||(P^TM)^{-1}||$ thereby mitigating the effect of overfitting \cite{chaturantabutnonlinear}, accumulation of errors can still take place as we will see during the numerical experiments. This accumulation can eventually result in severe errors and even numerical instability. To prevent this issue we propose regularizing the least-squares problem using a generalized Tikhonov regularization \cite{hastieelements}. Using such a regularization procedure we can penalize undesired properties of solutions to the DEIM optimization problem by adding extra terms to its associated cost function. In this article we will propose to use  regularization terms of the form:
\begin{equation}
    \boxed{\boldsymbol{c}(\boldsymbol{a}) = \argmin_{\widetilde{\boldsymbol{c}} \in \mathbb{R}^m} \left|\left|P^T C_h(\Phi \boldsymbol{a}) - P^T M \widetilde{\boldsymbol{c}} \right|\right|^2 + \alpha \left|\left|\widetilde{\boldsymbol{c}} - \boldsymbol{\mu}\right|\right|_\Gamma^2 \quad \text{s.t.} \quad \boldsymbol{a}^T \Phi^T M \widetilde{\boldsymbol{c}} = 0.}
    \label{eq:l2lsdeim}
\end{equation}
Here $||\boldsymbol{c}||_{\Gamma} := \sqrt{\left<\boldsymbol{c}, \Gamma \boldsymbol{c}\right>}$ with $\Gamma \in \mathbb{R}^{m\times m}$ symmetric positive definite (SPD), $\boldsymbol{\mu} \in \mathbb{R}^m$ is some desired or reference state and $\alpha \in \mathbb{R}^+$ is a hyperparameter. The minimization problem can again be solved using the method of Lagrange multipliers. The Lagrangian $\mathcal{L}(\boldsymbol{c},\lambda;\alpha) : \mathbb{R}^m \times \mathbb{R} \rightarrow \mathbb{R}$ of this minimization is defined as:
\begin{equation*}
    \mathcal{L}(\boldsymbol{c},\lambda;\alpha) = \left|\left|P^TC_h(\Phi \boldsymbol{a}) - P^T M \boldsymbol{c} \right|\right|^2 + \alpha \left|\left|\boldsymbol{c} - \boldsymbol{\mu}\right|\right|_\Gamma^2 + \lambda \boldsymbol{a}^T\Phi^TM\boldsymbol{c},
\end{equation*}
where $\lambda \in \mathbb{R}$ is a Lagrange multiplier. Taking partial derivatives of the Lagrangian and setting them to zero leads to the following system for the optima:
\begin{equation}
    \begin{bmatrix}
                   2 \left((P^T M)^T P^T M + \alpha \Gamma\right) &  (\boldsymbol{a}^T \Phi^T M)^T \\
                   \boldsymbol{a}^T \Phi^T M & 0 
    \end{bmatrix}
    \begin{bmatrix}
                   \boldsymbol{c}_o \\
                   \lambda_o
    \end{bmatrix} =
    \begin{bmatrix}
                   2(P^T M)^T P^T C_h(\Phi \boldsymbol{a}) + 2\alpha\Gamma \boldsymbol{\mu}\\
                   0
    \end{bmatrix}.
    \label{eq:lagrsys}
\end{equation}
Again, this matrix is symmetric and has constant coefficients with the exception of the last row and column which depend on $\boldsymbol{a}$. Denoting:
\begin{equation*}
    A_\Gamma := (P^T M)^T P^T M + \alpha \Gamma \in \mathbb{R}^{m \times m},
\end{equation*}
the DEIM coordinates as calculated by the regularized CLSDEIM are given as:
\begin{equation}
\addtolength{\fboxsep}{3pt}
\boxed{
\begin{gathered}
    \text{Regularized CLSDEIM: }\\
    \boldsymbol{c}(\boldsymbol{a}) = A_{\Gamma}^{-1}\left((P^T M)^{T} P^T C_h(\Phi \boldsymbol{a}) + \alpha\Gamma \boldsymbol{\mu}\right) - \frac{\boldsymbol{b}(\boldsymbol{a})^T A_{\Gamma}^{-1}\left((P^T M)^{T} P^T C_h(\Phi \boldsymbol{a}) + \alpha\Gamma \boldsymbol{\mu}\right)}{(\boldsymbol{b}(\boldsymbol{a})^T A_\Gamma^{-1} \boldsymbol{b}(\boldsymbol{a}))} A_{\Gamma}^{-1} \boldsymbol{b}(\boldsymbol{a}).
\end{gathered}
}\label{eq:l2clsdeim}
\end{equation}
Note that this equation also applies to a combination of oversampling and regularization. Taking $\alpha = 0$ in equation \eqref{eq:l2clsdeim} we obtain the normal CLSDEIM since $A_{\Gamma}^{-1}(P^T M)^T = (P^T M)^{\dagger} = (P^T M)^{-1}$ in this case. A data-driven method to calculate $\Gamma$ and $\boldsymbol{\mu}$ that uses the knowledge we have on the DEIM coordinates of our snapshot data $\Xi$ will be proposed in the following section. The hyperparameter tuning of $\alpha$ was carried out manually for simplicity. A more advanced  approach to find well-performing hyperparameters for regularized minimization problems like \eqref{eq:l2lsdeim} is generalized cross-validation \cite{golubgeneralized}, which we suggest for future work.

\subsubsection{Mahalanobis distance}
The method we propose to determine the regularization term is to use the Mahalanobis distance $d_M(\boldsymbol{c}) : \mathbb{R}^m \rightarrow \mathbb{R}^+$ \cite{hastieelements} of the solution $\boldsymbol{c}$ with respect to the data matrix $C = M^T \Xi \in \mathbb{R}^{m\times n_s}$, note the difference with the discrete convection operator $C_h$. The Mahalanobis distance is computed as:
\begin{equation*}
    d_M(\boldsymbol{c}) = \left|\left|\boldsymbol{c} - \boldsymbol{\mu}_C\right|\right|_{S^{-1}},
\end{equation*}
and thus corresponds to taking $\Gamma = S^{-1}$ and $\boldsymbol{\mu} = \boldsymbol{\mu}_C$ in optimization problem \eqref{eq:l2lsdeim}. Here, the matrix $S \in \mathbb{R}^{m \times m}$ and vector $\boldsymbol{\mu}_C \in \mathbb{R}^m$ are the covariance matrix and the mean associated to the data in $C$, respectively. This method is based on the idea of promoting solutions $\boldsymbol{c}$ that are similar to those observed in the data matrix $C$ (which has the DEIM coordinates of the snapshots in $\Xi$ as its columns). Here, similar is in the sense of originating from the same distribution. In fact, regularization using the Mahalanobis distance $d_M$ corresponds to assuming a prior distribution on the optimal $\boldsymbol{c}$ that can be accurately characterized by a mean and a covariance matrix $S$. The Mahalanobis distance is then the multidimensional generalization of the number of standard deviations an observation is removed from the distribution's mean. 

In this paper we will concern ourselves with solution reproduction problems for simplicity, and not with parameterized problems. We will see that, as numerical errors accumulate during time integration, DEIM coordinates take increasingly different values from their reference values in $C$ when using the DEIM or CLSDEIM. Hence, such values will contribute to a large $d_M(\boldsymbol{c})$ and as a result be less optimal than a more well-behaved solution given the regularized CLSDEIM loss-function \eqref{eq:l2lsdeim}. We assume that this idea can be extended to the parametric case, i.e.\ that the DEIM coordinates of a parametric problem can also be treated as generated by a single prior distribution. We will refer to the CLSDEIM regularized by the Mahalanobis distance as the MCLSDEIM.

The diagonal entries $(S)_{ii}$ of $S$ can be very small for $i \gg 1$, due to the fact that the DEIM coordinates for these $i$ are generally very small. Furthermore $S$ may be nearly rank deficient. Both these aspects can result in an ill-conditioned $S^{-1}$ having large entries. Future research will involve improving the conditioning of the regularization term using e.g.\ a penalized Mahalanobis distance \cite{hastieelements}. In the current work we solve this issue by taking very small hyperparameter values $\alpha$.

\section{A structure-preserving temporal localization method} \label{sec:time}
Having constructed new energy-conserving and robust DEIM methods, we now extend the feasibility of our approach with a new structure-preserving temporal localization method. Namely, it is well known that as flows become more convection dominated ($\nu\rightarrow 0$) the Kolmogorov N-width decay becomes increasingly slower. This makes it more difficult for conventional POD to generate a low-dimensional basis set that accurately captures a significant amount of energy contained in both $\Xi$ and the solution snapshots $X \in \mathbb{R}^{N \times n_s}$. To deal with this problem researchers have proposed several solutions, one of which is the application of the principal interval decomposition (PID) for both the construction of DEIM and POD spaces, $\mathcal{M}_d$ and $\mathcal{V}$ respectively \cite{chatarantabuttemporal, peherstorferlocalized, signalijzerman, intervalboggaard, ahmedbreaking, grimbergmesh, copelandreduced}. However, as far as we know structure-preservation has not been taken into account in existing PID approaches. In what follows, we will propose a method to preserve structure throughout a full time integration using a hROM with temporal localization.

\subsection{Temporal localization using PID} 
The PID decomposes the snapshot sets over $n$ intervals in time $[t_i, t_{i+1}]$ and applies the POD algorithm to the individual intervals. The idea is that by calculating modes tailored to specific intervals, the local timescales within the respective intervals are captured significantly better than by a set of modes calculated from the full set of snapshots \cite{intervalboggaard, chatarantabuttemporal}. Based on snapshot sets:
\begin{equation*}
    X = [X_1, X_2, ..., X_{n}], \quad \Xi = [\Xi_1, \Xi_2, ..., \Xi_{n}],
\end{equation*}
the PID provides sets of POD modes:
\begin{equation}
    \Phi_i, M_i, \hbox{ } i \in \{1,...,n\}, \label{eq:bases}
\end{equation}
applicable to use at times $t \in [t_i, t_{i+1}]$ within their respective intervals. The local POD modes $\Phi_i$ follow from an SVD of the local snapshot matrix $X_{i} \in \mathbb{R}^{N \times n_s^i}$, where $n_s^i \in \mathbb{N}$ is the number of snapshots in the $i^{\text{th}}$ interval. The temporally localized DEIM measurement space $\mathcal{P}_i$ for the $i^{\text{th}}$ interval is determined solely based on the DEIM modes in $M_i$ using the algorithm as in \cite{chaturantabutnonlinear} specifically for $M_i$. Furthermore, we have $\mathcal{V}^i := \text{span}(\Phi_i) \subset \mathbb{R}^N$ and $\mathcal{M}_d^i := \text{span}(M_i) \subset \mathbb{R}^N$. 

Setting up the hROM using the PID, the dynamical system \eqref{eq:deimrom} takes the form:
\begin{equation}
    \frac{d\boldsymbol{a}_i}{dt} = - \Phi_i^T M_i \boldsymbol{c}_i(\boldsymbol{a}_i) + \nu D_r^i \boldsymbol{a}_i \quad t \in [t_i, t_{i+1}],
    \label{eq:rnspiddeim}
\end{equation}
where $D_r^i := \Phi_i^T D_h \Phi_i$. Note that we also introduce subscripts for the generalized and DEIM coordinates $\boldsymbol{a}_i$ and $\boldsymbol{c}_i$ respectively, as they are only valid during the interval $[t_i, t_{i+1}]$. The DEIM coordinates are calculated using any of the previously discussed methods using the appropriate measurement spaces and DEIM and POD bases. It should be noted that all the conservation properties of the structure-preserving DEIM methods hold \textit{within} intervals. This is trivial as nothing changes in the way the DEIM coordinates are calculated. Instead, the art is to develop interface conditions such that during the transition from one interval to the next, these conservation properties are still satisfied. 

Note that in the subsequent sections a hROM using the PID and a certain DEIM algorithm will be referred to as PID-CLSDEIM hROM (when for example the CLSDEIM is used without regularization).

\subsection{Interface conditions}
The bases in \eqref{eq:bases} are used only during the interval $t\in [t_i, t_{i+1}]$. At $t = t_{i+1}$ an interface is reached and it is required to transition from a solution described in terms of $\boldsymbol{u}_r^- := \Phi_i\boldsymbol{a}_i(t_{i+1}) \in \mathcal{V}^{i}$ to one described in terms of $\boldsymbol{u}_r^+ := \Phi_{i+1} \boldsymbol{a}_{i+1}(t_{i+1}) \in \mathcal{V}^{i+1}$, where $\boldsymbol{a}_i \in \mathbb{R}^{r_i}$, $\boldsymbol{a}_{i+1} \in \mathbb{R}^{r_{i+1}}$ and in general $\text{dim}(\mathcal{V}^i) = r_i \neq r_{i+1} = \text{dim}(\mathcal{V}^{i+1})$. During the numerical experiments in this article we will consider the case $r_i = r \hbox{ } \forall \hbox{ } i \in \{1,...,n\}$. To determine the interval parameters, one could use heuristic methods  \cite{chatarantabuttemporal} or a clustering algorithm \cite{grimbergmesh}, but this is outside the scope of this article. 

We will define how to transition between intervals based on an interface condition. Using the conventional PID \cite{ahmedbreaking} the following interface condition is defined:
\begin{equation}
    \text{find} \quad \boldsymbol{a}_{i+1}(t_{i+1}) \quad \text{s.t.} \quad  \left<\boldsymbol{u}_r^+ - \boldsymbol{u}_r^-, \boldsymbol{v}\right>_{\Omega_h} = 0  \quad \forall \hbox{ } \boldsymbol{v} \in \mathcal{V}^{i+1}, \label{eq:interfacecond1}
\end{equation}
which is stable in the sense that it does not lead to an energy increase, as we will show. This condition is satisfied by the $\Omega_h$-orthogonal projection of $\boldsymbol{u}_r^-$ onto $\mathcal{V}^{i+1}$:
\begin{equation}
    \boldsymbol{a}_{i+1}(t_{i+1}) = \Phi_{i+1}^T \Omega_h \boldsymbol{u}_r^- = \Phi_{i+1}^T \Omega_h \Phi_i \boldsymbol{a}_{i}(t_{i+1}). \label{eq:intsol1}
\end{equation}
Note that the interface matrix $T_i := \Phi_{i+1}^T \Omega_h \Phi_i \in \mathbb{R}^{r_{i+1} \times r_i}$ can be precomputed in an offline stage to obtain a cheap online evaluation. The kinetic energy $K_r^+ := \frac{1}{2}\left|\left| \boldsymbol{u}_r^+\right|\right|_{\Omega_h}^2 = \sum_k\frac{1}{2}\left<\boldsymbol{u}_r^-,(\Phi_{i+1})_{,k}\right>_{\Omega_h}^2 $ cannot increase beyond $K_r^- := \frac{1}{2}\left|\left| \boldsymbol{u}_r^-\right|\right|_{\Omega_h}^2$ due to the Bessel inequality \cite{kreyszigintroductory}, which states that for the (complete) inner product space defined as $\mathcal{H}_{\Omega_h} := \left(\mathbb{R}^N, \left<\cdot,\cdot\right>_{\Omega_h}\right)$ and orthonormal sequence $\left((\Phi_{i+1})_{,k}\right)$ with $(\Phi_{i+1})_{,k} \in \mathcal{H}_{\Omega_h}, \hbox{ } k \in \{1,...,r_{i+1}\}$ we have the following:
\begin{equation*}
    \text{Bessel Inequality in $\mathcal{H}_{\Omega_h}$:} \quad \sum_{k=1}^{r_{i+1}}\left<\boldsymbol{u}_r^-,(\Phi_{i+1})_{,k}\right>_{\Omega_h}^2 \leq \left|\left| \boldsymbol{u}_r^-\right|\right|_{\Omega_h}^2.
\end{equation*}
Hence, application of the conventional PID using interface condition \eqref{eq:interfacecond1} cannot lead to an increase in kinetic energy for any number of intervals $n$, making it \textit{nonlinearly stable}. Furthermore, the solution is also optimal in the sense that \eqref{eq:intsol1} is also the solution to the minimization problem:
\begin{equation}
    \boldsymbol{a}_{i+1}(t_{i+1}) = \argmin_{\boldsymbol{a}\in \mathbb{R}^{r_{i+1}}} \left|\left| \boldsymbol{u}_r^- - \Phi_{i+1}\boldsymbol{a} \right| \right|_{\Omega_h}^2.
    \label{eq:intoptim1}
\end{equation}
 In summary, the conventional PID is nonlinearly stable. However, kinetic energy and momentum are in general not exactly conserved when transitioning between intervals using condition \eqref{eq:interfacecond1}, which can lead to artificial (numerical) dissipation of the solution. To solve this issue of lack of conservation, we propose the following new structure-preserving interface condition.

\subsection{Structure-preserving interface conditions}
\begin{figure}
    \centering
    \begin{minipage}{0.45\textwidth}
        \centering
        \includegraphics[width=1\textwidth]{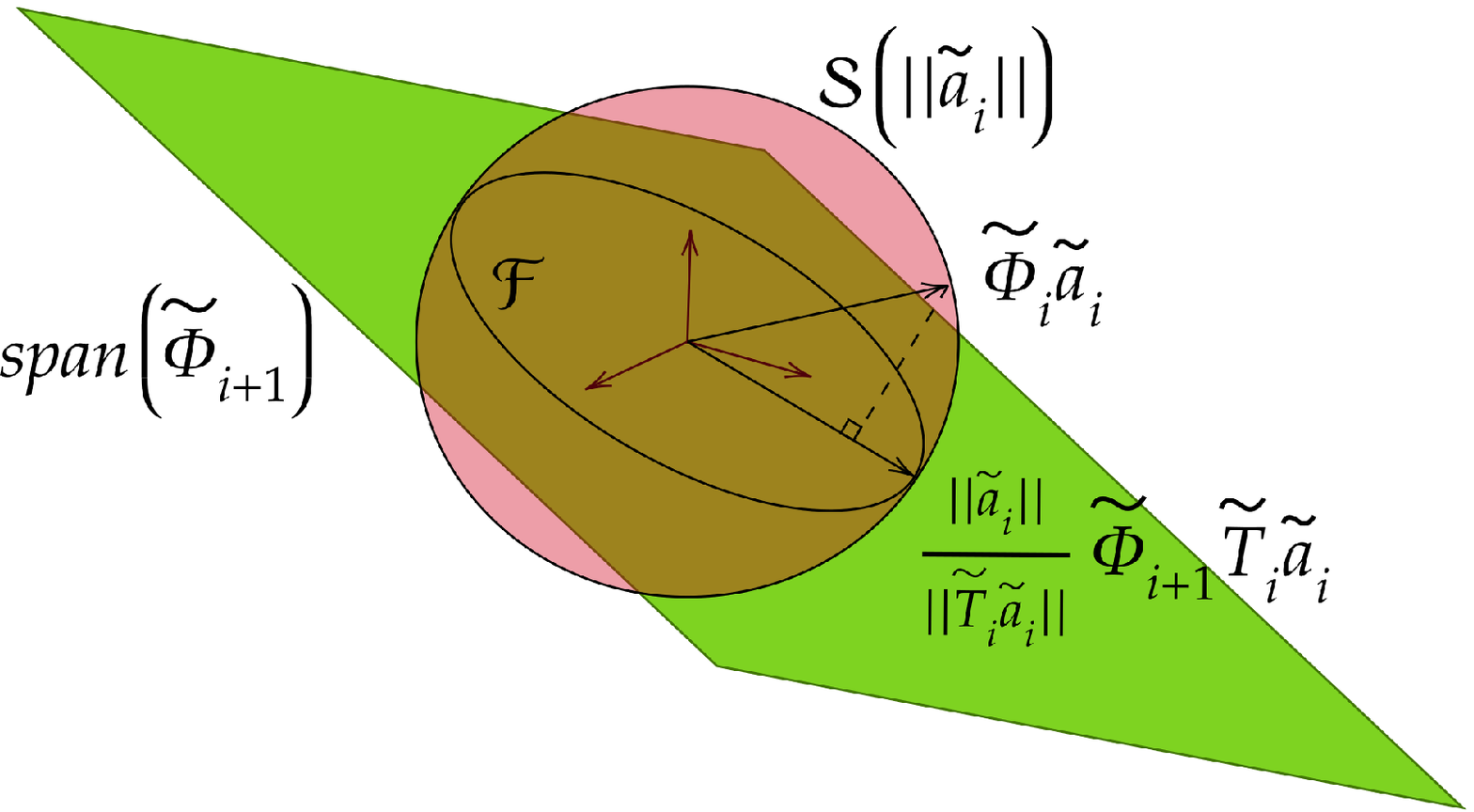} 
        \caption{Graphical intuition for optimization solution with $N = 3$ and $\Omega_h = I$.}
        \label{fig:optsol}
    \end{minipage}\hfill
    \begin{minipage}{0.45\textwidth}
        \centering
        \includegraphics[width=1\textwidth]{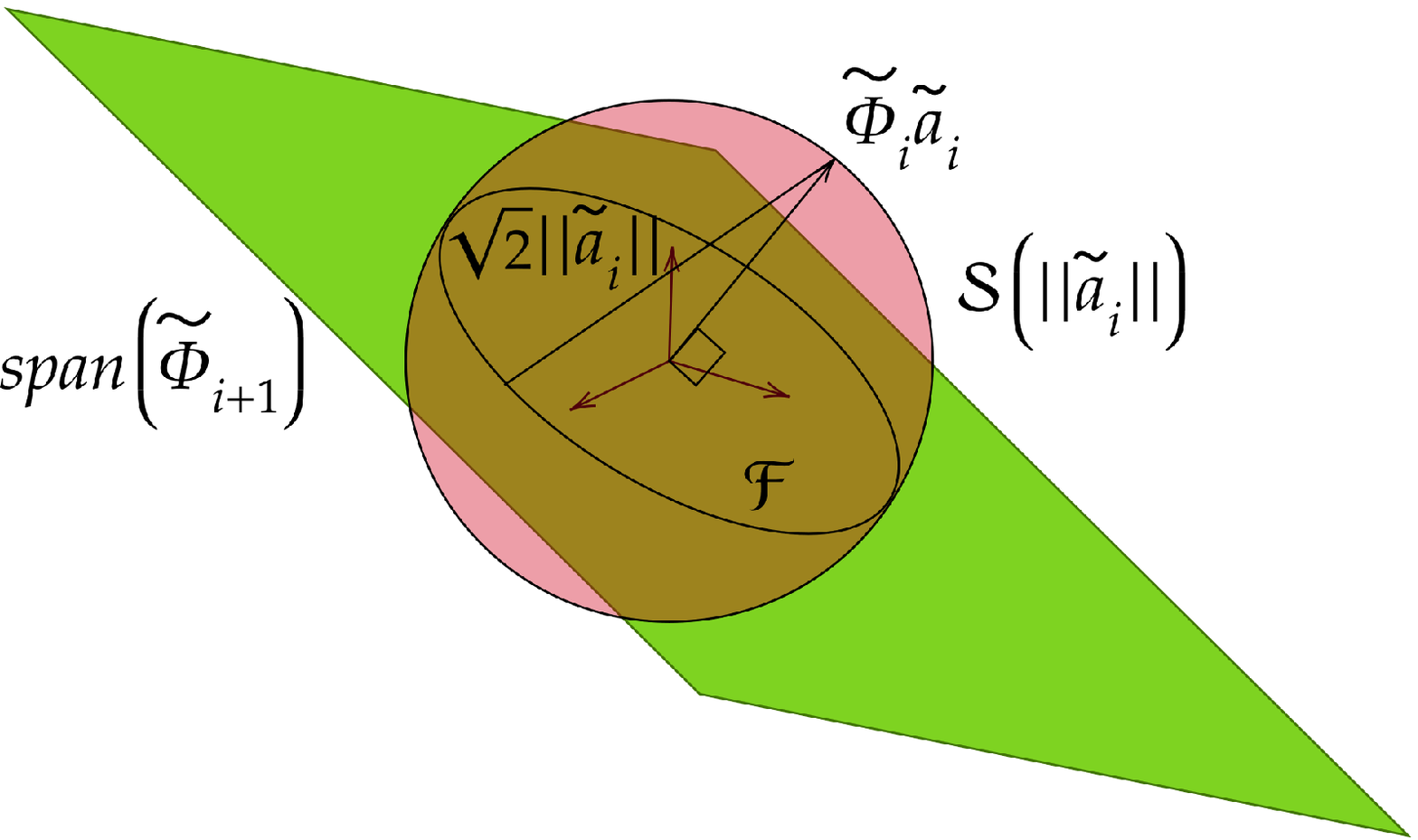} 
        \caption{Graphical intuition for infinite number of optimization solutions with $N = 3$ and $\Omega_h = I$}
        \label{fig:infoptsol}
    \end{minipage}
\end{figure}

To consider structure preservation over interfaces we define $\left(\boldsymbol{P}_r^+\right)_k := \left<\boldsymbol{e}_k, \boldsymbol{u}_r^+\right>_{\Omega_h}$ and similarly $\left(\boldsymbol{P}_r^-\right)_k := \left<\boldsymbol{e}_k, \boldsymbol{u}_r^-\right>_{\Omega_h}$. In order to conserve energy and momentum over interfaces we benefit from the optimization formulation \eqref{eq:intoptim1} underlying the non-conserving interface condition \eqref{eq:interfacecond1}. We propose to constrain the optimization problem \eqref{eq:intoptim1} to conserve $K_r$ and $\boldsymbol{P}_r$, leading to:
\begin{equation}
    \boxed{\boldsymbol{a}_{i+1}(t_{i+1}) = \argmin_{\boldsymbol{a}\in \mathbb{R}^{r_{i+1}}} \left|\left| \boldsymbol{u}_r^- - \Phi_{i+1}\boldsymbol{a} \right| \right|_{\Omega_h}^2 \quad \text{s.t.} \quad \frac{1}{2}\left|\left|\boldsymbol{a}\right|\right|^2 = K_r^-, \quad \left<\boldsymbol{e}_k, \Phi_{i+1}\boldsymbol{a}\right>_{\Omega_h} = \left(\boldsymbol{P}_r^-\right)_k \forall k \in \{1,...,d\}.}
    \label{eq:intoptim}
\end{equation}
The constraints on this problem can be simplified by realizing that the momentum component $(\boldsymbol{P}_r)_k$ in direction $k$ is fully determined by the $k^{\text{th}}$ component $a_{k}$ of the generalized coordinates, namely:
\begin{align*}
    \left(\boldsymbol{P}_r\right)_k = \left<\boldsymbol{e}_k,\sum_{j=1}^{r} \Phi_{,j} a_j\right>_{\Omega_h} = \left<\boldsymbol{e}_k, \frac{\boldsymbol{e}_k}{\left|\left|\boldsymbol{e}_k\right|\right|_{\Omega_h}}a_{k}\right>_{\Omega_h} = \left|\left|\boldsymbol{e}_k\right|\right|_{\Omega_h} a_{k},
\end{align*}
due to $\Omega_h$-orthogonality. Furthermore, the first $d$ POD modes for every interval are the same and given by the momentum-conserving modes $\boldsymbol{e}_k/\left|\left|\boldsymbol{e}_k\right|\right|_{\Omega_h}$. As a result, the momentum-conserving interface constraint can only be satisfied by setting $(\boldsymbol{a}_{i+1})_k = (\boldsymbol{a}_i)_k \hbox{ } \forall k \in \{1,...,d\}$. The following matrices and vectors are now defined:
\begin{equation*}
    \widetilde{\Phi}_i := \left[ (\Phi_i)_{,d+1}, \hbox{ } ... \hbox{ } , \hbox{} (\Phi_i)_{,r_i} \right], \quad \widetilde{\boldsymbol{a}}_i^T := \left[(\boldsymbol{a}_i)_{d+1}, \hbox{ } ... \hbox{ } , \hbox{ }(\boldsymbol{a}_i)_{r_i} \right].
\end{equation*} 
Since the momentum conservation constraint fully determines the first $d$ components of the generalized coordinates $\boldsymbol{a}_{i+1}$ we can simplify the optimization problem \eqref{eq:intoptim} by formulating it in terms of the new $\widetilde{(\cdot )}$ variables. Note that at any interface:
\begin{align*}
    \boldsymbol{u}_r^{+} - \boldsymbol{u}_r^{-} &= \sum_{k = d+1}^{r_{i+1}} (\Phi_{i+1})_{,k}(\boldsymbol{a}_{i+1})_k - \sum_{l = d+1}^{r_{i}} (\Phi_{i})_{,l} (\boldsymbol{a}_{i})_l + \sum_{j = 1}^d \frac{\boldsymbol{e}_j}{||\boldsymbol{e}_j||_{\Omega_h}} ((\boldsymbol{a}_{i+1})_j - (\boldsymbol{a}_{i})_j) \\
    &= \widetilde{\Phi}_{i+1} \widetilde{\boldsymbol{a}}_{i+1} - \widetilde{\Phi}_{i} \widetilde{\boldsymbol{a}}_{i} 
\end{align*}
due to the momentum-conserving interface conditions. The kinetic energy constraint can also be simplified to $||\widetilde{\boldsymbol{a}}_{i+1}||^2 = ||\widetilde{\boldsymbol{a}}_i||^2$ since the first $d$ components of the generalized coordinates $\boldsymbol{a}_{i+1}$ and $\boldsymbol{a}_i$ are the same. The reduced minimization problem for the remaining $r_{i+1}-d$ components of the generalized coordinates $\boldsymbol{a}_{i+1}(t_{i+1})$ then becomes:
\begin{equation}
    \widetilde{\boldsymbol{a}}_{i+1}(t_{i+1}) = \argmin_{\boldsymbol{a} \in \mathbb{R}^{r_{i+1} - d}}\left|\left|\widetilde{\Phi}_i\widetilde{\boldsymbol{a}}_i - \widetilde{\Phi}_{i+1}\boldsymbol{a} \right|\right|_{\Omega_h}^2 \quad \text{s.t.} \quad \left|\left|\boldsymbol{a}\right|\right|^2 = \left|\left|\widetilde{\boldsymbol{a}}_i\right|\right|^2. \label{eq:optimred}
\end{equation}
As before, we solve this constrained minimization problem using the method of Lagrange multipliers \cite{boydconvex}. The Lagrangian $\mathcal{L}(\boldsymbol{a},\lambda) : \mathbb{R}^{r_{i+1}-d} \times \mathbb{R} \rightarrow \mathbb{R}$ of this minimization is as follows:
\begin{equation*}
    \mathcal{L}(\boldsymbol{a}, \lambda) = \boldsymbol{a}^T\boldsymbol{a} - 2\boldsymbol{a}^T \widetilde{T}_i\widetilde{\boldsymbol{a}}_i + \lambda (\boldsymbol{a}^T \boldsymbol{a} - \widetilde{\boldsymbol{a}}_i^T\widetilde{\boldsymbol{a}}_i), 
\end{equation*}
where $\widetilde{T}_i := \widetilde{\Phi}_{i+1}^T \Omega_h \widetilde{\Phi}_i$ and $\lambda \in \mathbb{R}$ is a Lagrange multiplier. Taking partial derivatives of the Lagrangian and setting them to zero leads to the following nonlinear system for the optima ($\boldsymbol{a}_o, \lambda_o$):
\begin{gather*}
    \nabla_a \mathcal{L} = (1 + \lambda_o)\boldsymbol{a}_o - \widetilde{T}_i\widetilde{\boldsymbol{a}}_i = 0 \\
    \nabla_\lambda \mathcal{L} = \boldsymbol{a}_o^T \boldsymbol{a}_o - \widetilde{\boldsymbol{a}}_i^T\widetilde{\boldsymbol{a}}_i = 0.
\end{gather*}
Rewriting the equation in the first line, an expression for the generalized coordinates at the optimum $\boldsymbol{a}_o$ is obtained. This expression can be substituted into the second optimality condition to solve for $\frac{1}{1+\lambda_o}$. Finally, this results in:
\begin{equation}
    \widetilde{\boldsymbol{a}}_{i+1}(t_{i+1}) = \boldsymbol{a}_o =  \frac{||\widetilde{\boldsymbol{a}}_i||}{\left|\left|\widetilde{T}_i\widetilde{\boldsymbol{a}}_i\right|\right|} \widetilde{T}_i\widetilde{\boldsymbol{a}}_i. \label{eq:gencoordsoptim}
\end{equation}
To interpret this solution, we consider the subset $\mathcal{S}(\delta) := \{\boldsymbol{u}\in \mathcal{H}_{\Omega_h} \hbox{ } | \hbox{ } ||\boldsymbol{u}||_{\Omega_h} = \delta \}$ of all vectors in $\mathcal{H}_{\Omega_h}$ with the same norm of $\delta \in \mathbb{R}^+$ (and thus the same kinetic energy) and define the feasible set of minimization problem \eqref{eq:optimred}:
\begin{equation*}
    \mathcal{F}_a := \{ \boldsymbol{a} \in \mathbb{R}^{r_{i+1}-d} \hbox{ } | \hbox{ } \widetilde{\Phi}_{i+1} \boldsymbol{a} \in \mathcal{S}(||\widetilde{\boldsymbol{a}_i}||)\}
\end{equation*}
and the set $\mathcal{F} := \text{span}(\widetilde{\Phi}_{i+1}) \cap \mathcal{S}(||\widetilde{\boldsymbol{a}_i}||) = \{ \widetilde{\Phi}_{i+1} \boldsymbol{a} \hbox{ } | \hbox{ } \boldsymbol{a} \in \mathcal{F}_a\}$. The solution then corresponds to an orthogonal projection (in $\mathcal{H}_{\Omega_h}$) of $\widetilde{\Phi}_i \widetilde{\boldsymbol{a}}_i$ on $\text{span}(\widetilde{\Phi}_{i+1})$ and a subsequent rescaling to obtain an element of $\mathcal{F}$. Solution \eqref{eq:gencoordsoptim} then produces the generalized coordinates with respect to $\widetilde{\Phi}_{i+1}$ of this scaled projection. A visual interpretation for $N = 3$ and $\Omega_h = I$ is provided in \autoref{fig:optsol}.

Using equation \eqref{eq:gencoordsoptim} and the momentum conservation constraint the generalized coordinates of $\boldsymbol{u}_r^+$ as determined by the fully constrained minimization problem \eqref{eq:intoptim} are given by:
\begin{equation*}
    \boxed{\boldsymbol{a}_{i+1}(t_{i+1})^T = \left[
    (\boldsymbol{a}_i)_1 \hbox{ }, \hbox{ } ... \hbox{ } , \hbox{ }(\boldsymbol{a}_i)_d \hbox{ } , \left(\frac{||\widetilde{\boldsymbol{a}}_i||}{||\widetilde{T}_i\widetilde{\boldsymbol{a}}_i||} \widetilde{T}_i\widetilde{\boldsymbol{a}}_i\right)^T
    \right].}
\end{equation*}
We will refer to using the PID with this structure-preserving interface condition as SP-PID.

\begin{figure}
    \centering
    \includegraphics[width = 0.6\textwidth]{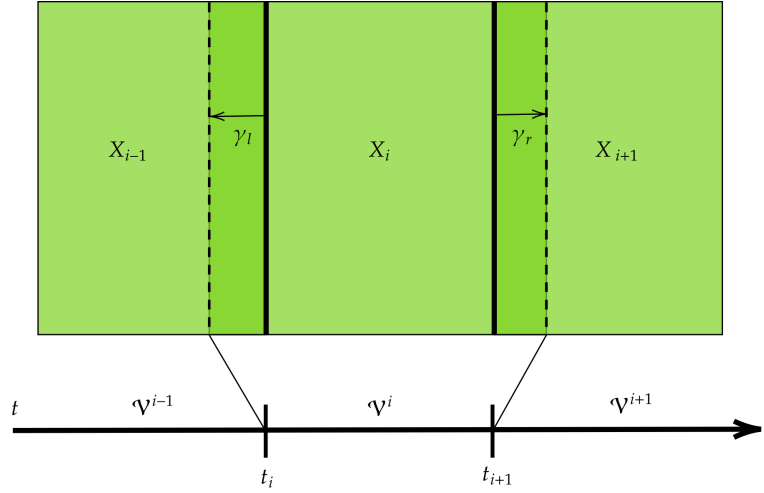}
    \caption{Constructing $\mathcal{V}^i$ using overlapping snapshots.}
    \label{fig:overlap}
\end{figure}

It may happen that $||\widetilde{T}_i\widetilde{\boldsymbol{a}}_i|| = 0$, in this case we cannot scale the result to have the same norm, and thus kinetic energy, as $\widetilde{\Phi}_i\widetilde{\boldsymbol{a}}_i$. This situation occurs when $\widetilde{\Phi}_i\widetilde{\boldsymbol{a}}_i \perp_{\Omega_h} \text{span}(\widetilde{\Phi}_{i+1})$ and it can be shown that, if $\text{dim}(\text{span}(\widetilde{\Phi}_{i+1})),N > 1$, minimization problem \eqref{eq:optimred} has an infinite number of solutions. Namely, writing out the loss function of minimization problem \eqref{eq:optimred}, taking into account that $\widetilde{\Phi}_i\widetilde{\boldsymbol{a}}_i \perp_{\Omega_h} \text{span}(\widetilde{\Phi}_{i+1})$, we have $||\widetilde{\Phi}_i\widetilde{\boldsymbol{a}}_i - \widetilde{\Phi}_{i+1}\boldsymbol{a}||_{\Omega_h} = \sqrt{2}||\widetilde{\boldsymbol{a}}_i|| \hbox{ } \forall \hbox{ } \boldsymbol{a} \in \mathcal{F}_a$ and since $|\mathcal{F}_a|$ is infinite in this case we have an infinite number of solutions. A visual representation of this case for $N = 3$ and $\Omega_h = I$ is given in \autoref{fig:infoptsol}. In principle any element of $\mathcal{F}_a$ can be used to determine $\boldsymbol{u}_r^+$ in this situation. We note that in general $||\widetilde{T}_i\widetilde{\boldsymbol{a}}_i|| = 0$ is an unlikely scenario since the snapshots at the end of $X_i$ tend to resemble those at the beginning of $X_{i+1}$. As a result, $\mathcal{V}^i$ and $\mathcal{V}^{i+1}$ are typically capable of resolving snapshots close to either side of their shared interface. However, to make sure that we prevent situations where $||\widetilde{T}_i\widetilde{\boldsymbol{a}}_i|| = 0$ (or close to zero) we suggest to use overlapping snapshots. This entails concatenating $\gamma_l \in \mathbb{N}$ of the rightmost snapshots in $X_{i-1}, \Xi_{i-1}$ and $\gamma_r \in \mathbb{N}$ of the leftmost snapshots in $X_{i+1}, \Xi_{i+1}$ with $X_i$ and $\Xi_i$ respectively for the construction of $\Phi_i$ and $M_i$, as illustrated in \autoref{fig:overlap}. This will promote non-empty and larger intersections between the reduced spaces and as a result discourage $||\widetilde{T}_i\widetilde{\boldsymbol{a}}_i|| = 0$, because both reduced spaces $\mathcal{V}^i$ and $\mathcal{V}^{i+1}$ are constructed to resolve $\boldsymbol{u}_r$ on either side of the interface.  In addition, the overlapping of snapshot sets tends to improve the accuracy of the transition, as will be shown in the results section.

\section{Results} \label{sec:results}
In this section we will discuss the results of two test cases to be introduced in what follows. The source code for these experiments is a custom finite volume incompressible flow solver that has been written in the C++ programming language using the Armadillo linear algebra library \cite{sandersonarmadillo, sandersonauser} and the LIS library of iterative solvers for linear systems \cite{nishidaexperience}. All experiments in this paper have been carried out on a simple workstation with a single Intel(R) Core(TM) i7-6700HQ CPU.

\subsection{Shear layer roll-up}
\subsubsection{Problem set-up}
The first test case concerns the shear layer roll-up (SLR) \cite{minionperformance, ronnieknikker}, which is a flow on a double-periodic domain $[0,2\pi] \times [0,2\pi]$ with a central band of flow in a positive coordinate direction and neighbouring bands of flow in the opposite direction. The bands are joined with a thin region of strong velocity gradients and hence strong shear forces. The initial conditions are: 
\begin{equation*}
    u(x,y,0) = \begin{cases}
        \text{tanh}\left(\frac{y - \pi / 2}{\delta}\right), & y \leq \pi\\
        \text{tanh}\left(\frac{3\pi / 2 - y}{\delta}\right), & y > \pi
    \end{cases}, \quad 
    v(x,y,0) = \epsilon \sin (x),
\end{equation*}
where  $\delta = \frac{\pi}{15}$ determines the initial thickness of the shear layers and the parameter $\epsilon= 0.05$ determines the initial amplitude of an unstable perturbation in the second coordinate direction to trigger the so-called roll-up. 

Using the SLR the proposed hROMs will be tested for their structure-preserving capabilities, accuracy and computational performance. Structure preservation will be tested by considering the temporal evolution of errors in conserved quantities (momentum and kinetic energy) for the first four seconds of the inviscid SLR flow. Accuracy will be tested by considering two different errors, namely the error with respect to the FOM, which is defined as
\begin{equation}
    \epsilon_u(t) = \left| \left| \boldsymbol{u}_h(t) - \boldsymbol{u}_r(t) \right| \right|_{\Omega_h},
\end{equation}
and the best-approximation error 
\begin{equation}
    \epsilon_b(t) = \left| \left| (I - \Phi \Phi^T \Omega_h) \boldsymbol{u}_h(t) \right|\right|_{\Omega_h},
\end{equation}
which forms a lower bound to the accuracy that can be obtained using $r$ POD modes. Comparing $\epsilon_u$ and $\epsilon_b$ will give an indication on how close to optimal the hROM is. We will also analyse whether overfitting takes place by considering the temporal evolution of individual DEIM coordinates and their Mahalanobis distance $d_M(\boldsymbol{c})$. Finally, the computational performance will be tested by measuring execution times associated to the offline and online phases of the hROMs and comparing these to the FOM. 

Based on a grid convergence study it was concluded that a $256 \times 256$ grid and time step $\Delta t =0.01$ were sufficiently fine to have a reliable FOM reference result, and these settings will be used in the numerical experiments unless otherwise mentioned. The same time step is also used to build the snapshot matrix and to simulate the ROMs. The hyper-parameter of the MCLSDEIM will be set to $\alpha = 1\cdot10^{-14}$ and the hyper-parameter of the OCLSDEIM will be set to $m_p = 2m$.

\begin{figure}[h]
    \centering
        \includegraphics[width=0.5\textwidth]{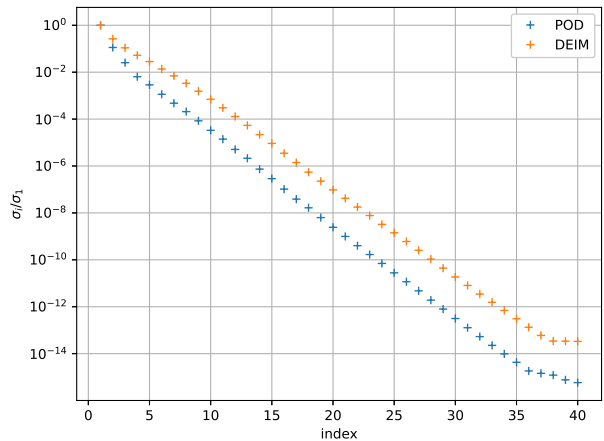} 
        \caption{The normalized POD and DEIM singular value decay of snapshot data from a FOM simulation with $\nu = 0$ and $t = 4$.}
        \label{fig:pod_svals_inv}
\end{figure}

\subsubsection{Conservation properties} 
The proposed hyper-reduction methods will be tested for their structure-preserving properties. We expect $K_r$ to be conserved for the inviscid case only when using the proposed hROMs in conjunction with energy-conserving Runge-Kutta (RK) methods (see remark \autoref{rmk:rk}). Thus, the proposed methods and the conventional DEIM will be tested for the case $\nu = 0$ using the (implicit) energy-conserving Gauss-Legendre 4 (GL4) method \cite{sanderseenergy}, and compared also to the classical (explicit) Runge-Kutta 4 (RK4) method \cite{sanderseaccuracy}. Time integration will take place until $t = 4$, since for longer time intervals numerical oscillations develop in the inviscid FOM simulation. These numerical oscillations do not destabilize the FOM solution due to the nonlinear stability property, but they render the simulation results inaccurate and corrupt the snapshot data sets and hence the ROM. Until $t = 4$ the FOM solution has a quickly decaying Kolmogorov N-width, and 8 POD and 8 DEIM modes suffice to accurately capture most of the energy in the snapshots as can be seen in \autoref{fig:pod_svals_inv}. Note that the POD and DEIM modes are obtained from snapshots taken at every individual time step of the FOM simulation. Since the initial conditions of the SLR have a net total momentum of zero in all coordinate direction we will plot the absolute value of its components $|(\boldsymbol{P}_r)_i|$. We expect this value to remain zero up to machine precision for all hROMs. For a momentum-conserving semi-discretization any RK method trivially conserves momentum, hence we only use RK4 (and not GL4) in order to not clutter the results. For the kinetic energy we will plot the value:
\begin{equation*}
    \epsilon_K(t) := |K_r(t) - K_r(0)|,
\end{equation*}
which denotes the absolute deviation of $K_r$ with respect to its initial value. As kinetic energy is conserved for $\nu  = 0$ by the structure-preserving hROMs, this value is also expected to remain zero up to machine precision using energy-conserving RK methods.

\begin{figure}
    \centering
    \begin{minipage}{0.45\textwidth}
        \centering
        \includegraphics[width=1\textwidth]{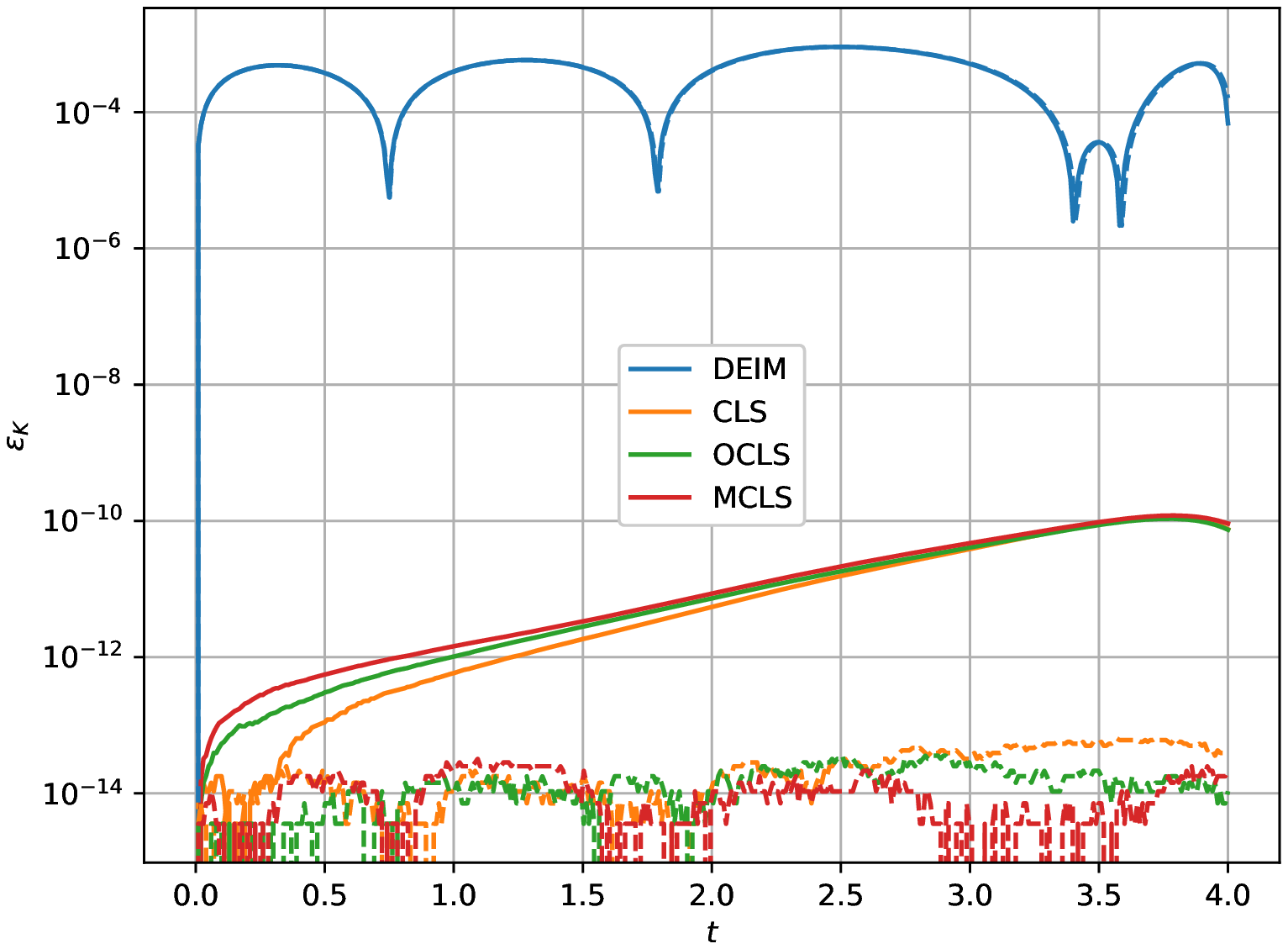} 
        \caption{The error $\epsilon_K$ in kinetic energy with respect to its initial value for the inviscid SLR using the proposed hROMs and the DEIM-hROM, the solid lines are for RK4 and the dashed lines for GL4.}
        \label{fig:energy_consrv}
    \end{minipage}\hfill
    \begin{minipage}{0.45\textwidth}
        \centering
        \includegraphics[width=1\textwidth]{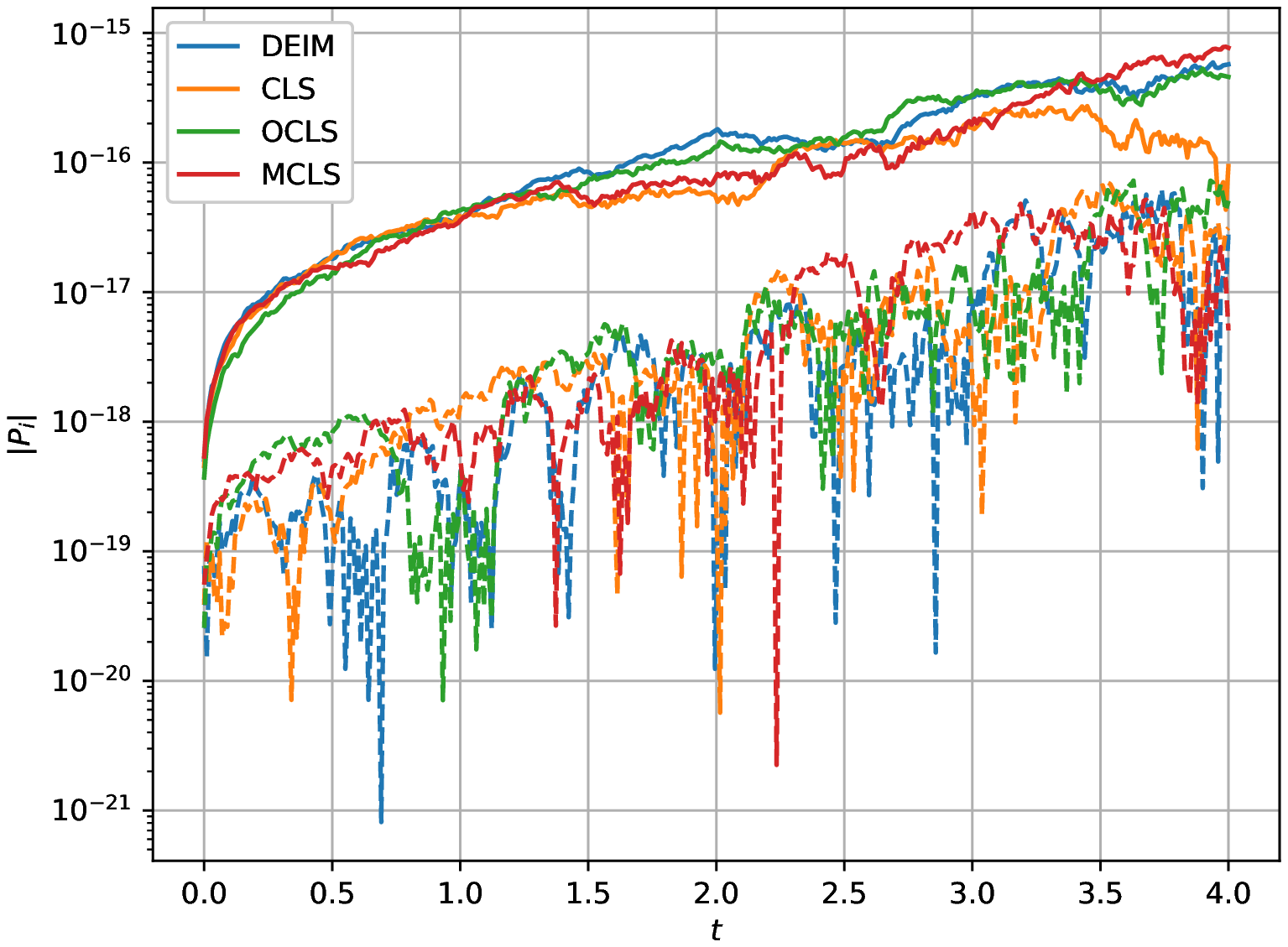} 
        \caption{Conservation of momentum component $(\boldsymbol{P}_r)_i$ in absolute value for the inviscid SLR using the proposed hROMs and the DEIM-hROM and RK4, the solid lines are for $i = 1$ and the dashed lines are for $i = 2$.}
        \label{fig:momentum_consrv}
    \end{minipage}
\end{figure}

In \autoref{fig:energy_consrv} the error in the kinetic energy is displayed. Colors are used to distinguish the different hROM methods and solid lines and dashed lines are used to distinguish RK4 from GL4 time integration. The key observation is that when using the standard DEIM algorithm (with either RK4 or GL4), $K_r$ deviates significantly from its initial value, while \textit{our proposed structure-preserving hROMs exactly conserve kinetic energy up to machine precision} (using GL4 time integration). When using the RK4 time integrator the structure-preserving methods deviate slightly from their initial value, due to the non-energy-conserving nature of RK4. These results are in line with \cite{sandersenonlinearly} in which  similar small energy errors were observed when using high-order explicit RK methods to integrate systems with structure-preserving semi-discretizations.

In \autoref{fig:momentum_consrv} the error in the momentum components is displayed. In this figure colors are used to distinguish which hROM method is applied and solid and dashed lines are used to indicate the first and second component of $\boldsymbol{P}_r$, respectively. It can be observed that the theoretical results on momentum conservation hold for all hROMs including the DEIM-hROM. The deviation from zero of all momentum components $|(\boldsymbol{P}_r)_i|$ can be seen to increase minutely. We assume this can be attributed to the finite precision of the SVD implementation of the Armadillo C++ library, which causes the modes in the DEIM basis $M$ to be momentum-conserving only up to the SVD precision.

\subsubsection{Convergence behaviour} 
\begin{figure}[h!]
    \centering
    \includegraphics[width=0.5\textwidth]{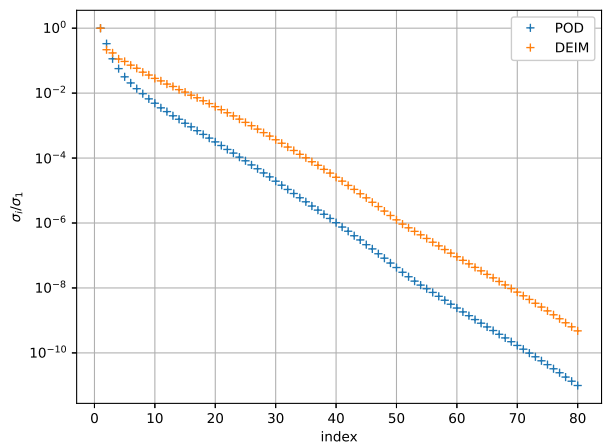} 
    \caption{The normalized POD and DEIM singular value decay of snapshot data from a FOM simulation with $\nu = 0.001$ and $t = 8$.}
    \label{fig:sval_pod_viscous}
\end{figure}

To assess the behaviour of $\epsilon_u$ as $r$ and $m$ are varied we perform a convergence study. This will be done for the case of viscous flow, with $\nu = 0.001$, allowing integration of the SLR flow until $t = 8$. Given the small differences between RK4 and GL4 in the study of conservation properties,  we only consider RK4 here for efficiency. Our primary interest is the behaviour of $\epsilon_u$ as a function of $m$, hence we will sweep this parameter for some values of $r$. To determine what values of $m$ and $r$ are necessary such that the corresponding reduced spaces $\mathcal{V}$ and $\mathcal{M}_d$ can accurately resolve the snapshot data we study again the singular value decay of $X$ and $\Xi$. The singular value decays of the snapshot data for a FOM simulation with $\nu = 0.001$ and until $t = 8$ have been plotted in \autoref{fig:sval_pod_viscous} for the data sets $X$ and $\Xi$. From this figure we determine a sweep will be performed over all values $m \in \{1, 2, ..., 60\}$ corresponding to a relative information content (RIC) \cite{lassilamodel} ranging from $89.86\%$ to $100\%$. The reduced space dimension will be taken as $r \in \{30,40\}$ both corresponding to $\text{RIC} \approx 100\%$.

\begin{figure}
    \centering
    \begin{minipage}{0.45\textwidth}
        \centering
        \includegraphics[width=1\textwidth]{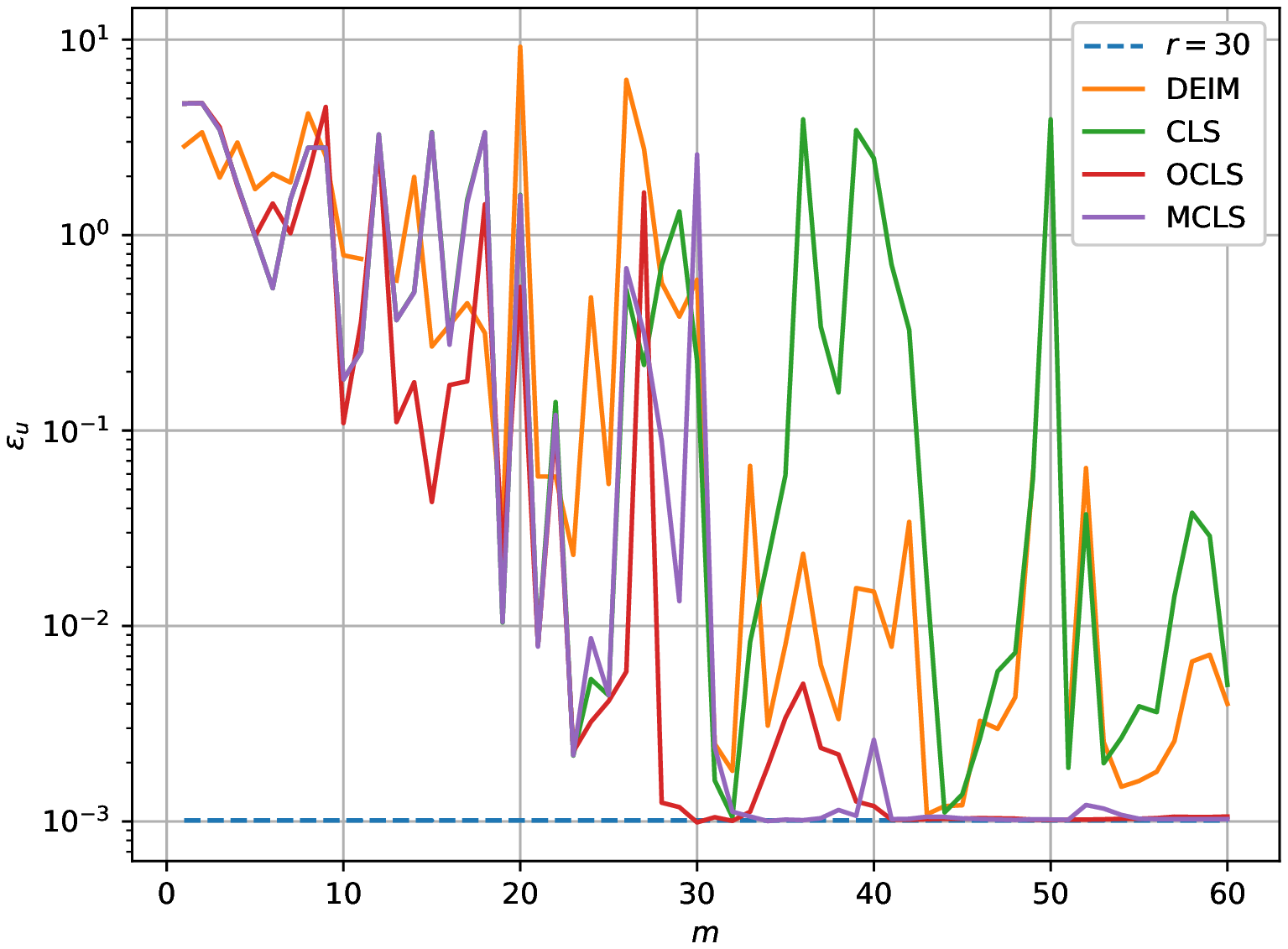} 
        \caption{Convergence study of $\epsilon_u$ evaluated at $t = 8$ for different hROMs and different values of $m$ for $r = 30$, including the non-hyper-reduced model results (dashed).}
        \label{fig:conv_r30}
    \end{minipage}\hfill
    \begin{minipage}{0.45\textwidth}
        \centering
        \includegraphics[width=1\textwidth]{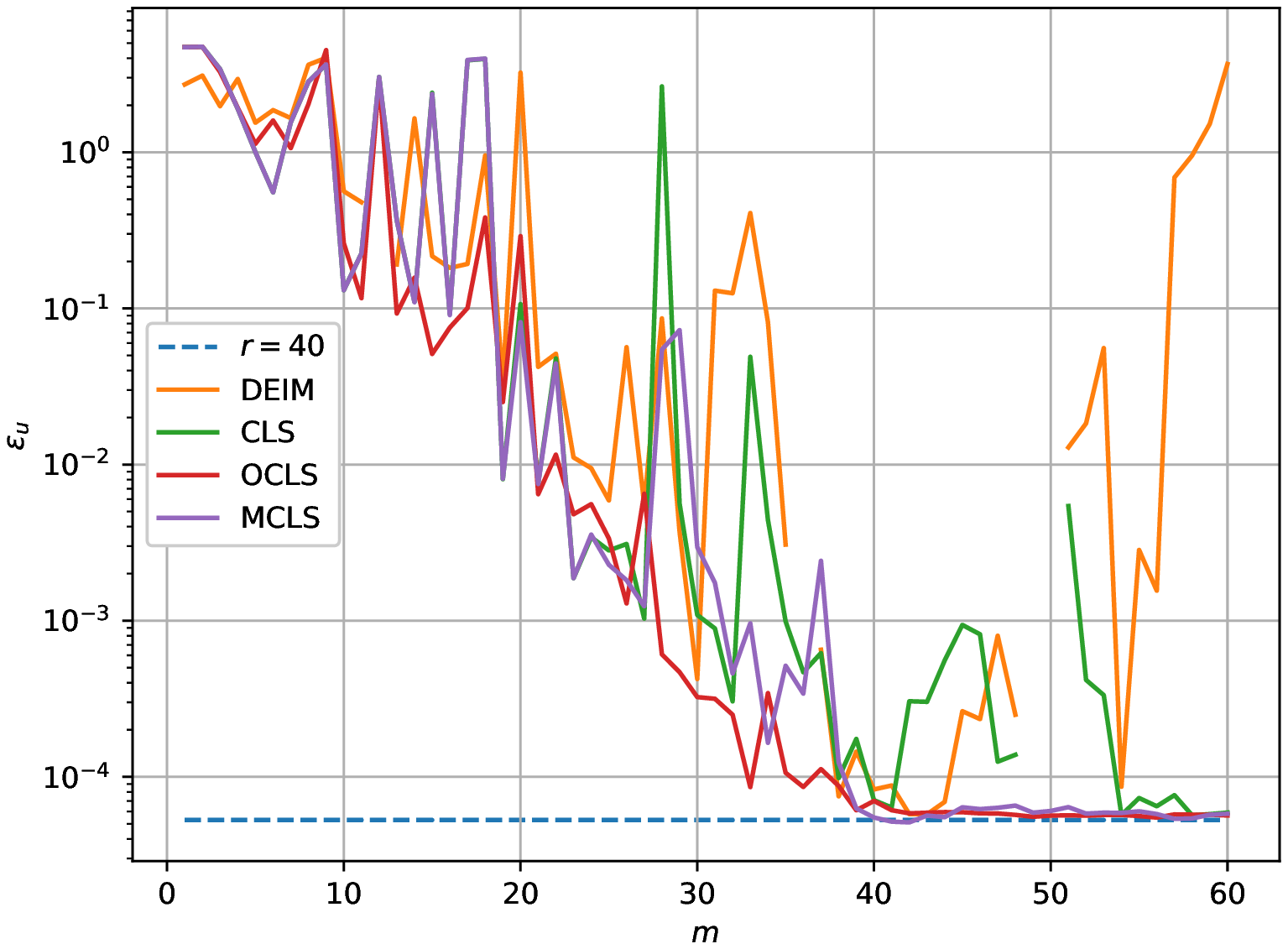} 
        \caption{Convergence study of $\epsilon_u$ evaluated at $t = 8$ for different hROMs and different values of $m$ for $r = 40$, including the non-hyper-reduced model results (dashed).}
        \label{fig:conv_r40}
    \end{minipage}
\end{figure}

The results of the convergence study have been plotted in \autoref{fig:conv_r30} and \autoref{fig:conv_r40} for $r = 30$ and $r = 40$ respectively. In these figures we also plot $\epsilon_u$ for the non-hyper-reduced order model (dashed lines). This value will be referred to as the `ROM error', obtained by using the exact tensor decomposition for the convective terms. Clearly, the ROM error is independent of the dimension $m$ of $\mathcal{M}_d$. Although the ROM error is not strictly a lower bound for $\epsilon_u$ of hROMs, the methods should converge to it if the convection operator is approximated accurately enough. All methods show erratic error behaviour as $m$ increases for $m < r$. This shows that the convection operator cannot be expected to be approximated well throughout a long time-integration procedure when $m$ is not sufficiently large compared to $r$. When $m \geq r$ better accuracy is obtained with the OCLSDEIM and MCLSDEIM, and the ROM error is closely approximated. This means that the convection operator is approximated well enough throughout the time integration such that $\Phi^T C_h(\Phi \boldsymbol{a})$ and $\Phi^T M\boldsymbol{c}$ are nearly indistinguishable. It can be seen that $\epsilon_u$ as obtained using the DEIM and CLSDEIM remains behaving erratically when $m \geq r$ and generally the ROM error is not obtained. This shows that exactly (DEIM) or near exactly (CLSDEIM) fitting the approximation $M\boldsymbol{c}$ to $C_h(\Phi \boldsymbol{a})$ in $\mathcal{P}$ when $m_p = m$ is not sufficient to guarantee accuracy in a long time integration process. Indeed, using more measurements (OCLSDEIM) or smarter choices of DEIM coordinates (MCLSDEIM) is required to obtain the ROM error. 

\begin{figure}
    \centering
    \includegraphics[width=\textwidth]{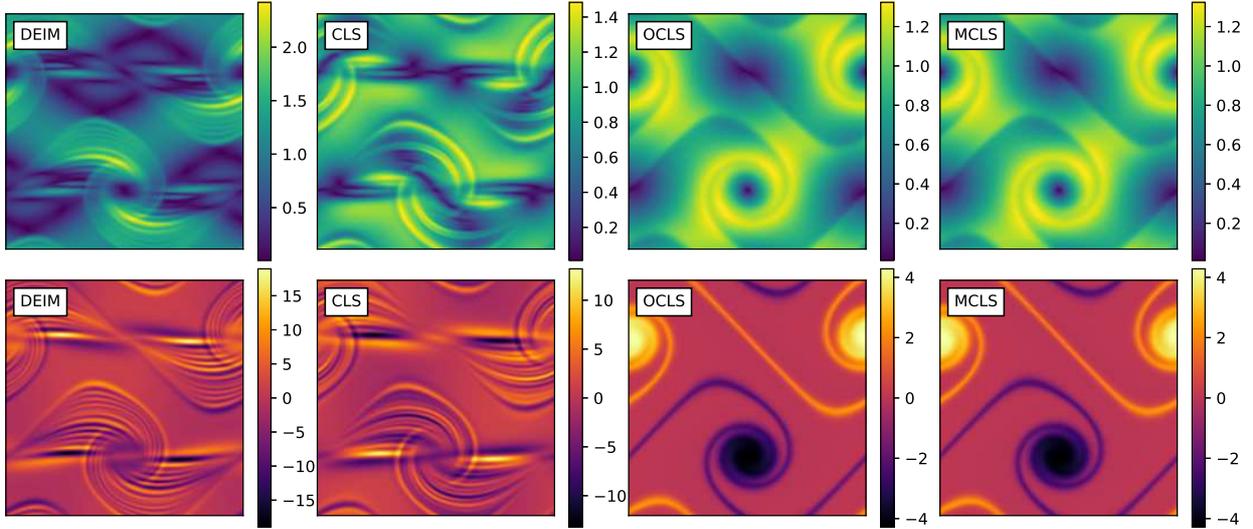}
    \caption{The velocity and vorticity fields as predicted by the different hyper-reduction methods at $r = 30, m = 50, t = 8$ ($t = 5$ for the unstable DEIM).}
    \label{fig:fieldsslr}
\end{figure}

\subsubsection{Temporal error evolution}
To obtain a better understanding of the performance of the methods we will analyse the full temporal evolution of several errors. We will examine a particularly bad case for the DEIM and CLSDEIM such that the errors are further exaggerated, specifically we will look at $r = 30, m = 50$, for which \autoref{fig:conv_r30} shows high errors for both DEIM and CLSDEIM. For illustration, the velocity and vorticity fields as obtained using $r=30, m=50$ are shown in \autoref{fig:fieldsslr}. For this case the error due to the DEIM becomes unstable and due to the CLSDEIM becomes very large. Besides $\epsilon_u$, we will also analyse the temporal evolution of the operator error:
\begin{equation*}
    \epsilon_M := \frac{||C_h(\Phi \boldsymbol{a}) - M\boldsymbol{c}(\boldsymbol{a})||}{||C_h(\Phi \boldsymbol{a})||}.
\end{equation*}
This error provides insight in how well a hyper-reduction method is approximating the convection operator.

\begin{figure}[h!]
\centering
\includegraphics[width=0.7\linewidth]{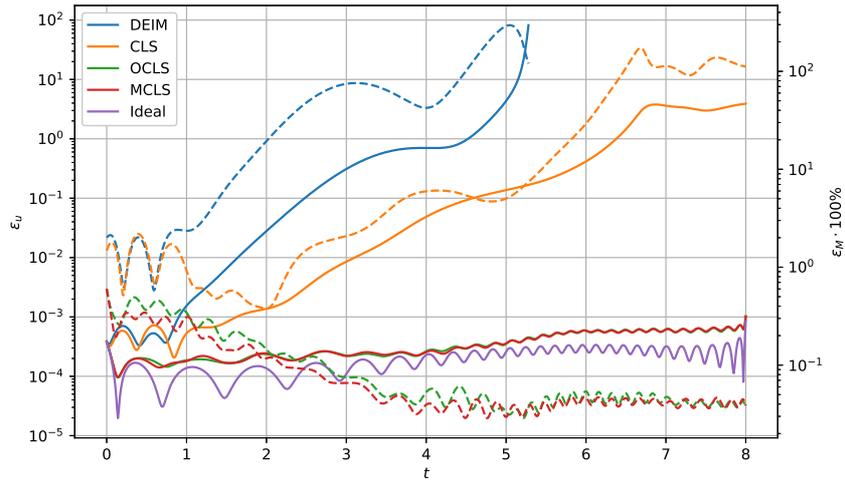}
\caption{The evolution of $\epsilon_u$ and $\epsilon_M$ (dashed lines) for $r=30$ and $m=50$ using different hROMs and for both RK4 (solid lines) compared to the ideal projection error $\epsilon_b$.}
\label{fig:Ng1} 
\end{figure}

In \autoref{fig:Ng1} the error evolution of both $\epsilon_u$ and $\epsilon_M$ have been displayed for $r = 30, m = 50$. The ideal projection error $\epsilon_b$ has also been plotted as function of time. We once more use colors to distinguish different methods, solid lines are associated to the evolution of $\epsilon_u$ using the RK4 integrator and dashed lines to $\epsilon_M$. We can see that the errors due to DEIM and CLSDEIM grow gradually as time progresses. Furthermore, as the solution becomes more dominated by errors it becomes more difficult for the DEIM and CLSDEIM to accurately approximate the correct convection operator output. The OCLSDEIM and MCLSDEIM are much more accurate throughout the full time integration process. It can be seen that their respective $\epsilon_u$ values stay close to the lower bound $\epsilon_b$ and that the operator output is approximated very accurately.

\begin{figure}[h!]
    \centering
    \includegraphics[width=1\textwidth]{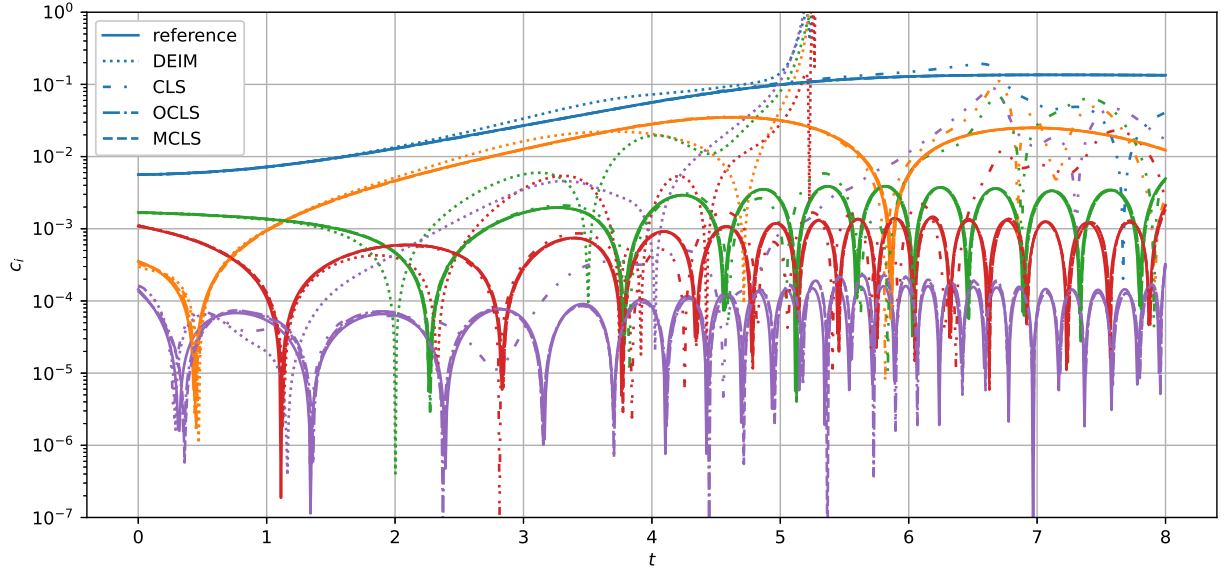} 
        \caption{Temporal evolution of DEIM coordinates $c_i$, $i \in (1,2,11,16,26)$ using $r = 30, m = 50$ compared to the reference values as given by the columns of $M^T\Xi$.}
        \label{fig:deimcoords}
\end{figure}

\begin{figure}[h!]
    \centering
    \includegraphics[width=0.5\textwidth]{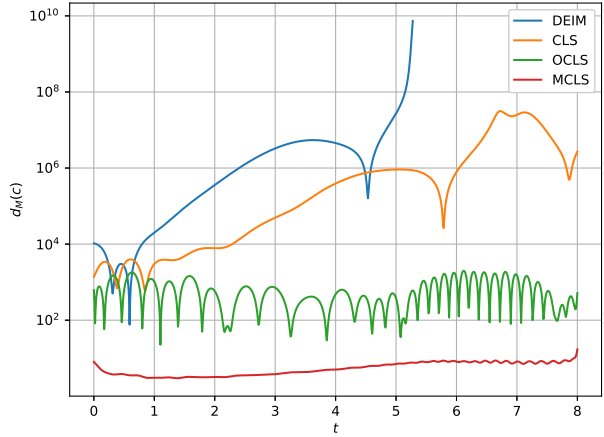} 
        \caption{Temporal evolution of the Mahalanobis distance $d_M(\boldsymbol{c})$ of the DEIM coordinates using $r = 30, m = 50$ for the different hyper-reduction methods.}
        \label{fig:mahalanobis}
\end{figure}

Another experiment that gives insight in the error is to plot the temporal evolution of individual DEIM coordinates. We have carried out this experiment again for the `bad' case $r = 30, m = 50$. The results are displayed in \autoref{fig:deimcoords}, where the evolution of a subset of the first 26 DEIM coordinates is given. Colors are used to indicate different DEIM coordinates and different linetypes are used to indicate what hyper-reduction method was used. We also plot the reference values for the DEIM coordinates as given by the columns of $M^T\Xi$. In this reproduction case the DEIM coordinates as given by the different hyper-reduction methods should be as close as possible to these ideal reference values. We can see the DEIM coordinates as determined by the DEIM gradually becoming unstable around 5 seconds into the simulation. The DEIM coordinates as determined by the CLSDEIM differ significantly from the reference values. The OCLSDEIM and MCLSDEIM perform much better and almost coincide with the reference values throughout the full simulation. Plotting the Mahalanobis distance $d_M(\boldsymbol{c})$ in \autoref{fig:mahalanobis}, the poor accuracy of the DEIM coordinates as determined by the DEIM and CLSDEIM is accentuated. As time progresses the values of the DEIM coordinates becomes increasingly less similar to the reference values in $M^T\Xi$, whereas the DEIM coordinates as determined by the OCLSDEIM and MCLSDEIM remain close to the reference distribution. Furthermore, \autoref{fig:mahalanobis} shows that the Mahalanobis distance, at least in the reproduction case, is a good indicator of a poor choice in DEIM coordinates as it becomes increasingly larger when $\epsilon_M$ becomes larger. Combining these results and the previous results on error evolution, we see that by exactly (DEIM) or near exactly (CLSDEIM) fitting the approximation $M\boldsymbol{c}$ to $C_h(\Phi \boldsymbol{a})$ in $\mathcal{P}$ with $m_p = m$, DEIM coordinates are determined that cause gradual accumulation of errors in the time integration process. Instead, more robustness is required to deal with demanding situation like time integration where accumulation of errors can cause problems. As shown in our experiments, robustness of the DEIM and CLSDEIM can be significantly improved by:
\begin{enumerate}
    \item performing a regression through a larger set of measurements than free parameters (OCLSDEIM);
    \item using knowledge on the prior distribution of DEIM coordinates (MCLSDEIM).
\end{enumerate}


\subsubsection{Performance}
To end our discussion on the SLR experiment we will show that the hyper-reduction methods exhibit similar computational performance as the conventional DEIM. This will be done by measuring the on- and offline phase wall-clock times and the associated speed-ups with respect to the FOM. Here, the offline phase is constituted by the necessary SVDs to construct $\mathcal{M}_d$ and $\mathcal{V}$, the construction of $\mathcal{P}$ and the precomputation of the necessary matrices and LU-decompositions. The online phase is constituted by the numerical integration of the hROMs. We will evaluate this for $r = 30, m = 40$ as all methods provide accurate results for this configuration of reduced space dimensions. To see the effect of increasing $m_p$ for the OCLSDEIM we will test for $m_p \in \{2m, 4m\}$, where we expect to see a decline in performance as $m_p$ is increased.

The results are tabulated in \autoref{tab:slronline} for the online and offline phase. It can be seen that the computational performance is similar for all methods, both in terms of online and offline performance. The DEIM, CLSDEIM and MCLSDEIM are especially close due to the fact that for all these cases $m_p = m$. The effect of the larger number of measurement evaluations required can be seen for the OCLSDEIM(2m) and OCLSDEIM(4m), which are slower than the other methods. Furthermore, using $m_p = 2m$ of the OCLSDEIM indeed provides better computational performance than $m_p = 4m$, as expected.

\begin{table}[h!]
\centering
\begin{tabular}{cccc} 
\hline
Method   & Wall-clock time online (s) & Speed-up & Wall-clock time offline (s)\\ \hline
DEIM     & 0.236               & 808      & 45.105 \\
CLSDEIM  & 0.300               & 636      & 45.104\\
OCLSDEIM(2m) & 0.494               & 386  & 44.972     \\
OCLSDEIM(4m) & 1.099               & 174  & 45.123    \\ 
MCLSDEIM & 0.302               & 632      & 44.215 \\ \hline
\end{tabular}
\caption{Computational performance (offline and online) of the different hyper-reduction methods for $r = 30, m = 40$.}
\label{tab:slronline}
\end{table}

\FloatBarrier

\subsection{Two-dimensional homogeneous isotropic turbulence}
\subsubsection{Problem set-up}
The second test case is two-dimensional homogeneous isotropic turbulence (2DT). 2DT is a flow on a double-periodic domain $[0,2\pi]\times[0,2\pi]$. We will generate initial conditions following the procedure outlined in \cite{omerhigh}. In short, we construct an initial energy spectrum:
\begin{equation*}
    E(k) = \frac{a_s}{2}\frac{1}{k_p}\left(\frac{k}{k_p}\right)^{2s+1} \exp{\left[-\left(s + \frac{1}{2}\right)\left(\frac{k}{k_p}\right)^2\right]},
\end{equation*}
where $k := ||\boldsymbol{k}|| = \sqrt{k_x^2 + k_y^2}$ is the wave-vector magnitude. The maximum value of the initial spectrum takes place at $k_p = 12$, $a_s \in \mathbb{R}$ is a normalization factor given by $a_s := (2s+1)^{s+1} / (2^ss!)$ and $s \in \mathbb{R}$ is a shape parameter taken as $s = 3$. From this spectrum we follow \cite{omerhigh} and calculate vorticity Fourier-coefficients like:
\begin{equation*}
    |\hat{\omega}(\boldsymbol{k})| = \sqrt{\frac{k}{\pi}E(k)}.
\end{equation*}
Consequently, we generate a vorticity spectrum $\hat{\omega}(\boldsymbol{k}) : \mathbb{R}^2 \rightarrow \mathbb{C}$ by providing random phases to the Fourier-coefficients using a process for which we refer to \cite{omerhigh}. Finally, we obtain a divergence-free velocity field by solving the relation:
\begin{equation*}
    \Delta \psi = -\omega, \quad \boldsymbol{u} = \begin{bmatrix}
        \frac{\partial \psi}{\partial y},
        -\frac{\partial \psi}{\partial x}
    \end{bmatrix}^T,
\end{equation*}
numerically. Note that this procedure is strictly for two-dimensional flows.

2DT is a fluid flow that has been studied extensively \cite{ahmedbreaking, davidsonturbulence, omerhigh}. Furthermore, it is a highly convection-dominated flow with many convected spatial structures and as a result it has a slowly decaying Kolmogorov N-width. For this reason the purpose of this test case is two-fold. Namely, the complexity of the flow makes it an adequate test case to see how well our proposed hyper-reduction methods will perform for more turbulent flows of engineering interest. Additionally, the slow Kolmogorov N-width decay will allow us to compare the PID and our newly proposed structure-preserving alternative, the SP-PID, in a meaningful setting.

The FOM simulations are performed with a $1024 \times 1024$ grid and a time step of $\Delta t = 0.0002$. These settings are based on the observation that these also led to converged results for similar spatially second-order energy-conserving schemes in \cite{omerhigh}. Saving snapshots every time step for this test
case is not feasible given the simple workstation used to perform all calculations. Hence, the simulation will be sampled such that the Nyquist-Shannon criterion is met \cite{ahmedbreaking}; this criterion is satisfied by sampling every $\Delta t_s = 0.005$.

\subsubsection{Hyper-reduction methods}
\begin{figure}[h!]
    \centering
    \begin{minipage}{0.45\textwidth}
        \centering
        \includegraphics[width=1\textwidth]{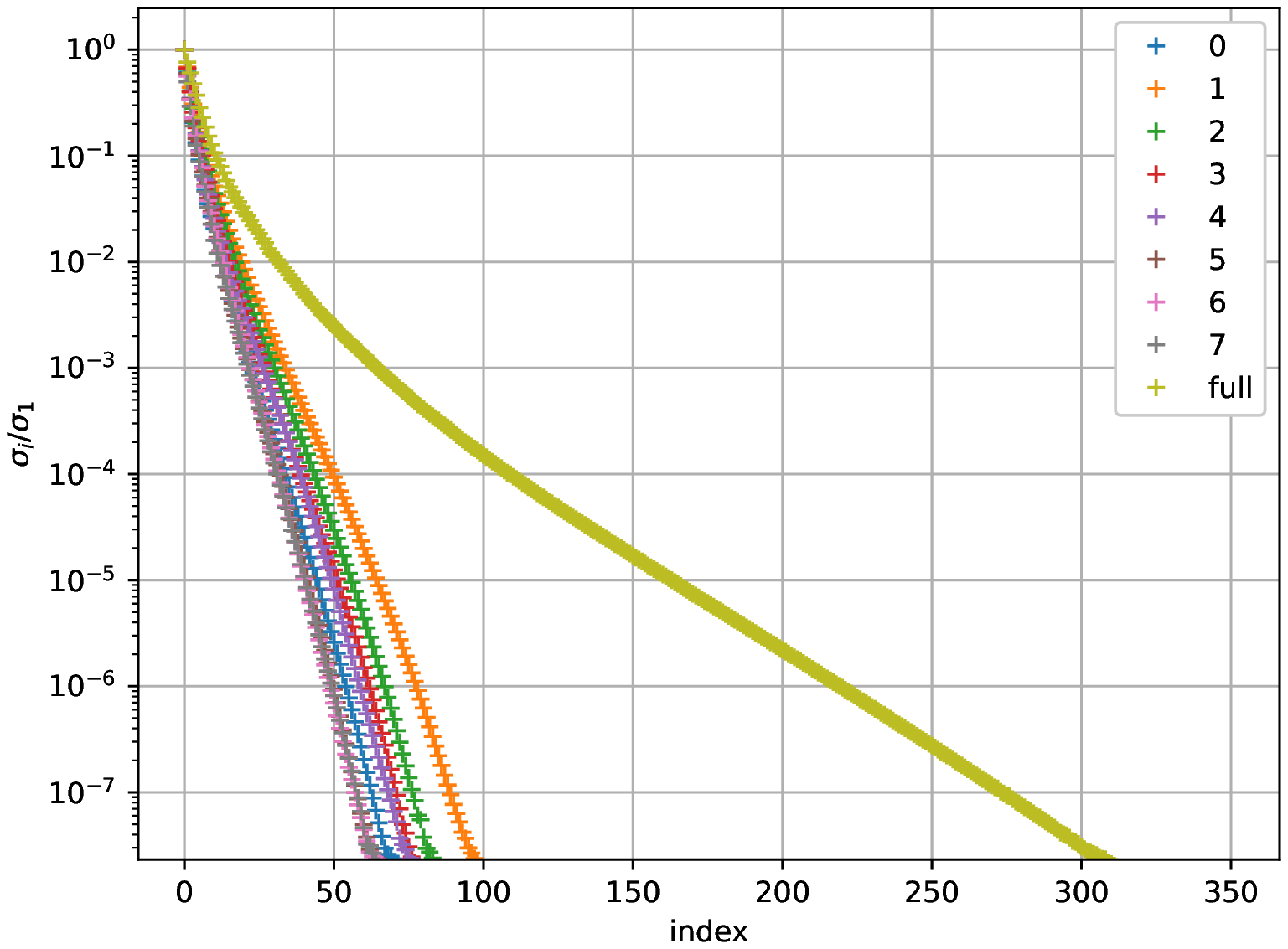} 
        \caption{Singular value decay of solution snapshots of intervals versus the full snapshot set.}
        \label{fig:pidpodsing}
    \end{minipage}\hfill
    \begin{minipage}{0.45\textwidth}
        \centering
        \includegraphics[width=1\textwidth]{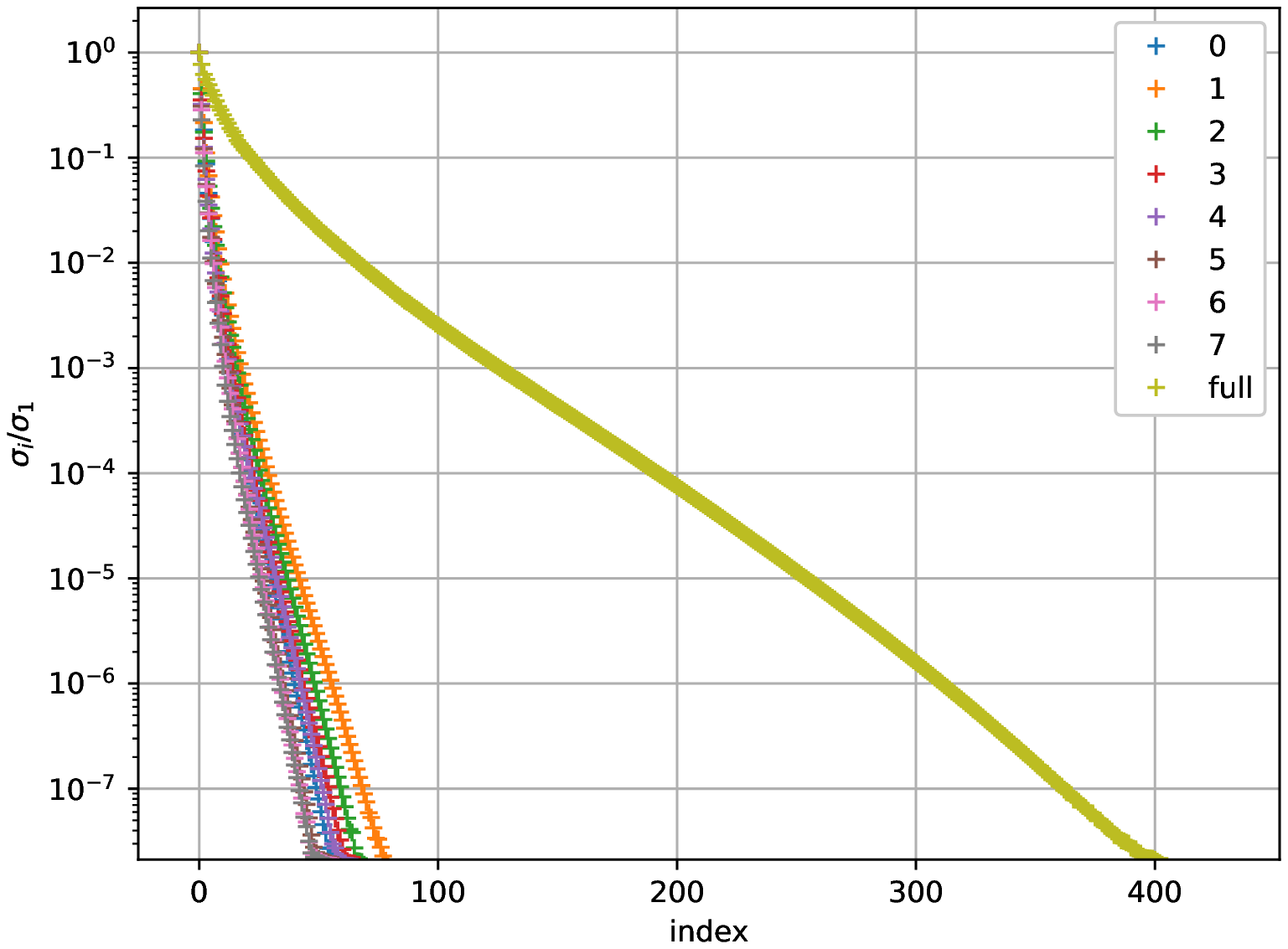} 
        \caption{Singular value decay of operator snapshots of intervals versus the full snapshot set.}
        \label{fig:piddeimsing}
    \end{minipage}
\end{figure}

 Due to the previous results for the SLR flow, in this section we will only test the two best-performing methods, MCLSDEIM and the OCLSDEIM. We simulate the FOM until $t = 4$ with $\nu = 0.001$ and compare how well the resulting energy spectra $E(k)$ of the FOM and hROMs align at $t \in \{1,2,3,4\}$. We will also provide a visual comparison of the vorticity fields at these time instances. To deal with the slow Kolmogorov N-width decay we will apply the SP-PID with $n = 8$ evenly sized intervals of 100 snapshots and for all intervals we will take an overlap parameters $\gamma_l = \gamma_r = \gamma = 20$. The singular values of all intervals and the full snapshot sets have been displayed in \autoref{fig:pidpodsing} and \autoref{fig:piddeimsing} for the POD and DEIM, respectively. As discussed in \cite{ahmedbreaking}, it can be seen that the application of temporal localization indeed leads to a faster decay of singular values within intervals. This effectively increases the Kolmogorov N-width decay. For the reduced space dimensions we will use $r = 30$ and $m = 40$ for all intervals. Like \cite{omerhigh}, we will calculate the energy spectra using the following equations:
 \begin{equation*}
     E(k,t) = \sum_{k \leq ||\widetilde{\boldsymbol{k}}|| < {k+1}} \hat{E}\left(\widetilde{\boldsymbol{k}},t\right), \quad \hat{E}(\boldsymbol{k},t) = \frac{1}{2} k^2 |\hat{\psi}(\boldsymbol{k},t)|^2,
 \end{equation*}
 where $\hat{\psi}(\boldsymbol{k},t) : \mathbb{R}^2 \times \mathbb{R}^+ \rightarrow \mathbb{C}$ are the Fourier-coefficients of the numerically calculated stream-function associated to a discrete velocity field $\boldsymbol{u}_h(t)$.

\begin{figure}[h!]
    \centering
    \begin{minipage}{0.45\textwidth}
        \centering
        \includegraphics[width=1\textwidth]{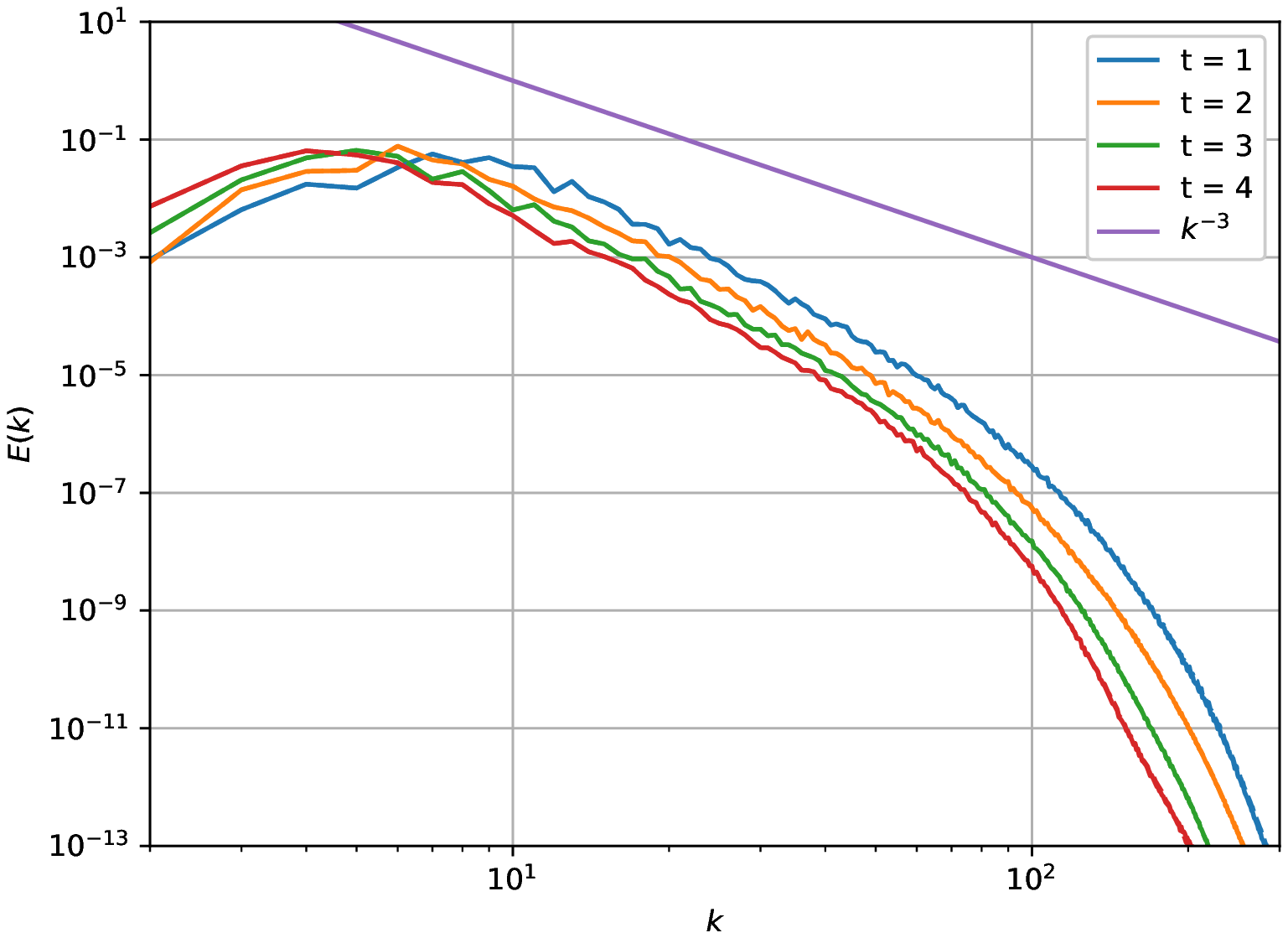} 
        \caption{Energy spectra at $t \in \{1,2,3,4\}$ as calculated by the FOM (solid line) and MCLSDEIM-SP-PID hROM (dashed line).}
        \label{fig:mclsspec}
    \end{minipage}\hfill
    \begin{minipage}{0.45\textwidth}
        \centering
        \includegraphics[width=1\textwidth]{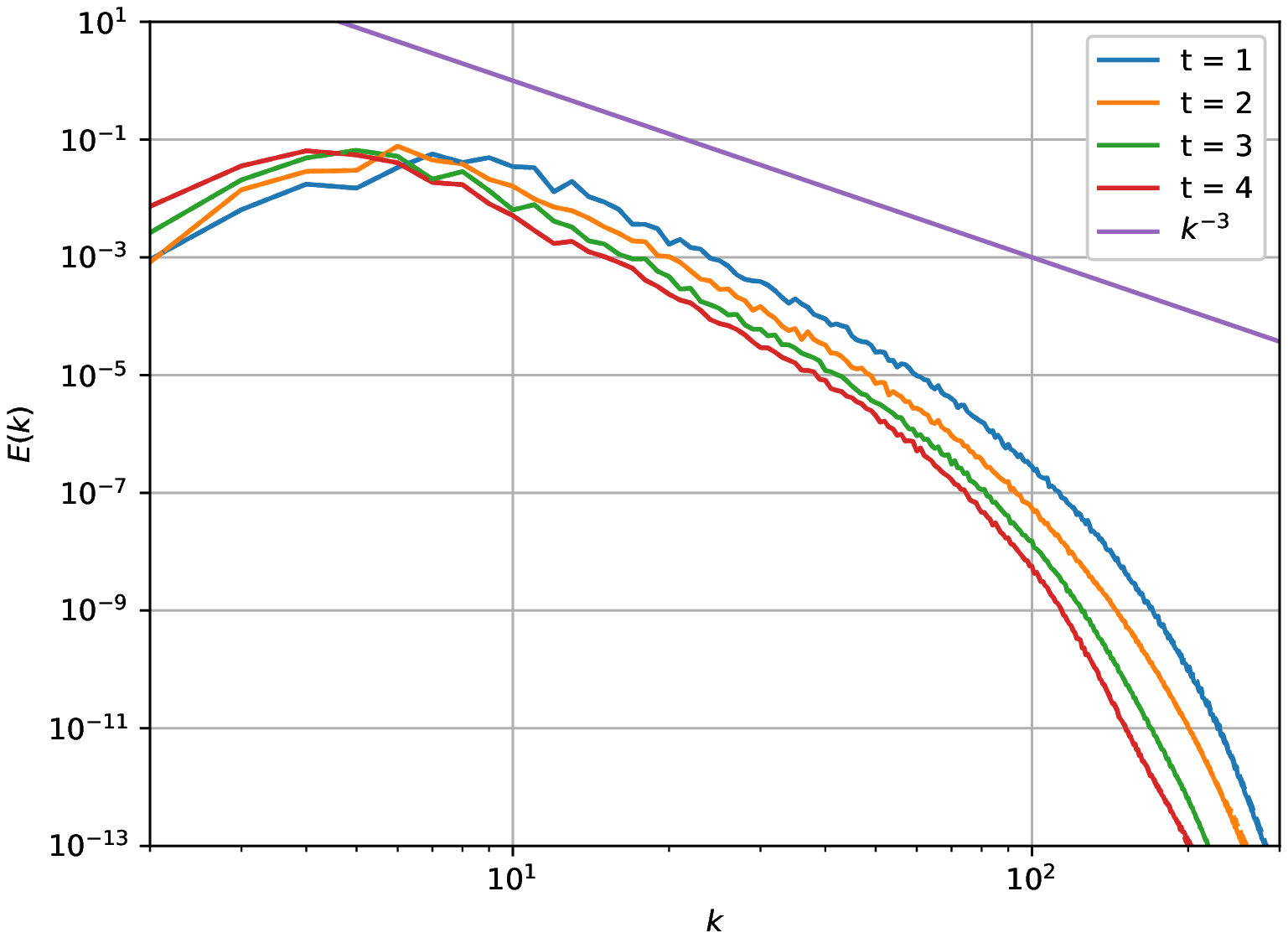} 
        \caption{Energy spectra at $t \in \{1,2,3,4\}$ as calculated by the FOM (solid line) and OCLSDEIM-SP-PID hROM (dashed line).}
        \label{fig:oclsspec}
    \end{minipage}
\end{figure}

\begin{figure}[b!]
    \centering
    \includegraphics[width = \textwidth]{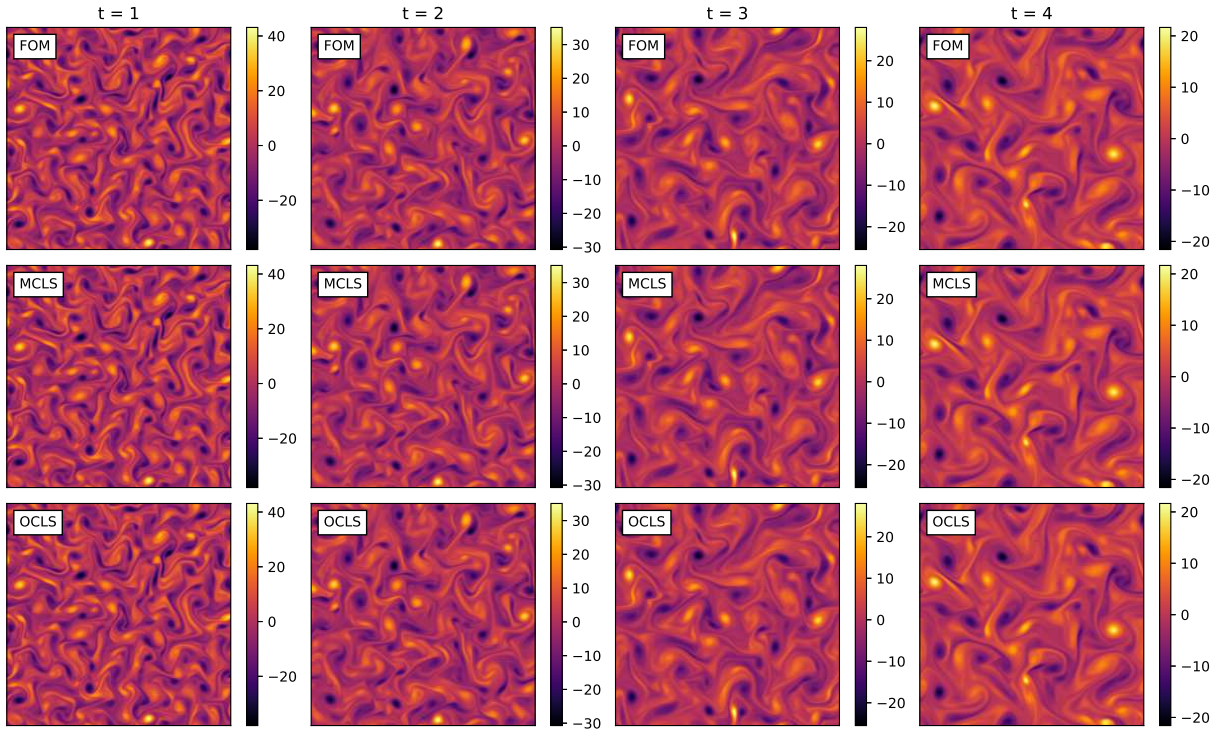}
    \caption{2DT vorticity fields as calculated using the FOM and hROMs for $t \in \{1,2,3,4\}$.}
    \label{fig:2dtvorts}
\end{figure}

We have provided the energy spectra for the different hyper-reduction methods in \autoref{fig:mclsspec} and \autoref{fig:oclsspec}. Excellent agreement at all points in time can be observed for both the hyper-reduction methods. Furthermore, the results are in line with those observed in \cite{omerhigh} for $Re = 1000$ in terms of spectra, where an accurate high-order pseudospectral method was used. We can see clearly that the spectrum is shifting to the left as time progresses, implying the presence of an inverse energy cascade. Furthermore, the inertial range nearly scales as $\mathcal{O}(k^{-3})$ which should be the case as $\nu \rightarrow 0$ \cite{davidsonturbulence}. We have also shown the vorticity fields as calculated using the FOM and the different hROMs in \autoref{fig:2dtvorts}. We can see that, besides the energy spectra matching between the FOM and hROMs, the vorticity fields are also almost identical. From this we conclude that the hROMs using the MCLSDEIM and OCLSDEIM and the SP-PID:
\begin{enumerate}
\item can accurately reconstruct the FOM results;
\item adhere to the physics underlying the 2DT flow.
\end{enumerate}


\subsubsection{Temporal localization methods}
We will now study in more detail the conservation properties of SP-PID at interfaces and compare its transition accuracy against the orthogonal projection of the standard PID as in equation \eqref{eq:intsol1}. Furthermore, we will briefly investigate the impact of the overlap parameter $\gamma$. We will do this by simulating the FOM until $t = 1$ with $\nu = 0.001$ and only considering $n = 2$ evenly sized intervals. We will then construct the reduced spaces associated to these intervals with overlap parameters $\gamma \in \{0,20\}$ and analyse the differences this makes on the accuracy of the transition at $t = 0.5$. We will also consider the accuracy of the conserved quantities energy and momentum, which should be exactly conserved by the SP-PID and not by the PID. However, for the PID, we expect these quantities to be better conserved for larger $\gamma$. We also expect the accuracy of the transition to be improved by taking larger $\gamma$, for both SP-PID and PID. During this experiment we will use the MCLSDEIM with $\alpha = 1\cdot10^{-14}$ and $m = 40$. Additionally, we will use a reduced space of $r = 30$. Finally, we will determine the accuracy of the transition using the transition error:
\begin{equation*}
    \epsilon_T := ||\boldsymbol{u}_r^+ - \boldsymbol{u}_r^-||_{\Omega_h}.
\end{equation*}

\begin{table}[h!]
\centering
\begin{tabular}{lllll}
\hline
    & SP-PID ($\gamma = 0$) & PID ($\gamma = 0$) & SP-PID ($\gamma = 20$) & PID ($\gamma = 20$) \\ \hline
$\epsilon_T$  & 0.0629     & 0.0629   & 0.00515     & 0.00515   \\
$|K_r^+ - K_r^-|$  & 0             & 0.00198  & 0              & $1.33\cdot 10^{-5}$  \\
$|(\boldsymbol{P}_r)_1^+ - (\boldsymbol{P}_r)_1^-|$ & 0             & $2.83\cdot 10^{-14}$ & 0              & $3.32\cdot 10^{-15}$  \\
$|(\boldsymbol{P}_r)_2^+ - (\boldsymbol{P}_r)_2^-|$ & 0             & $8.49\cdot 10^{-17}$ & 0              & $2.16\cdot 10^{-16}$  \\ \hline
\end{tabular}
\caption{Transition errors in velocity field and conserved quantities at the interface on $t = 0.5$.}
\label{tab:errpid}
\end{table}

\begin{figure}[h!]
    \centering
    \begin{minipage}{0.45\textwidth}
        \centering
        \includegraphics[width=1\textwidth]{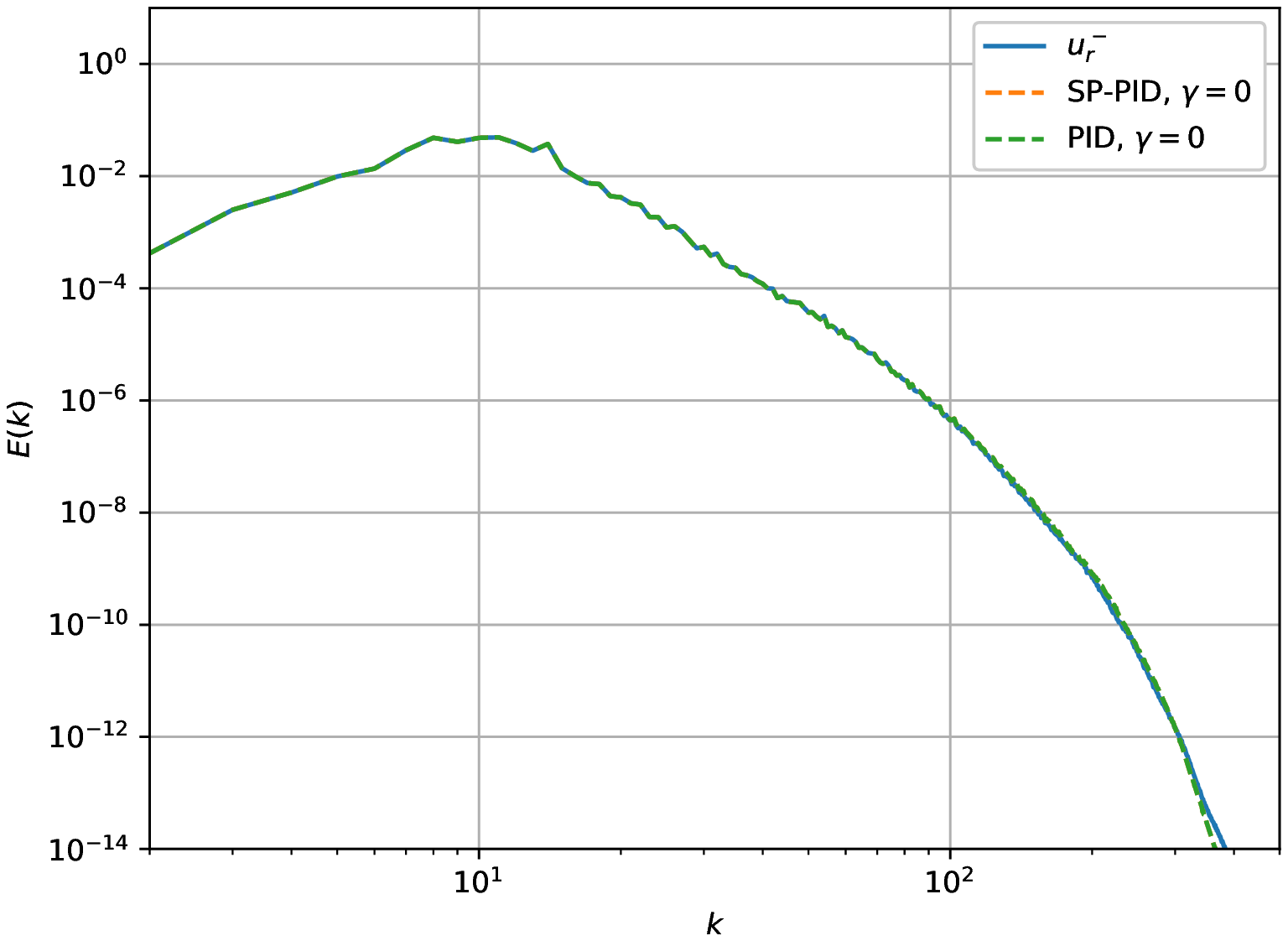} 
        \caption{Spectra before and after transition at $t = 0.5$ using PID and SP-PID for $\gamma = 0$.}
        \label{fig:umupom0}
    \end{minipage}\hfill
    \begin{minipage}{0.45\textwidth}
        \centering
        \includegraphics[width=1\textwidth]{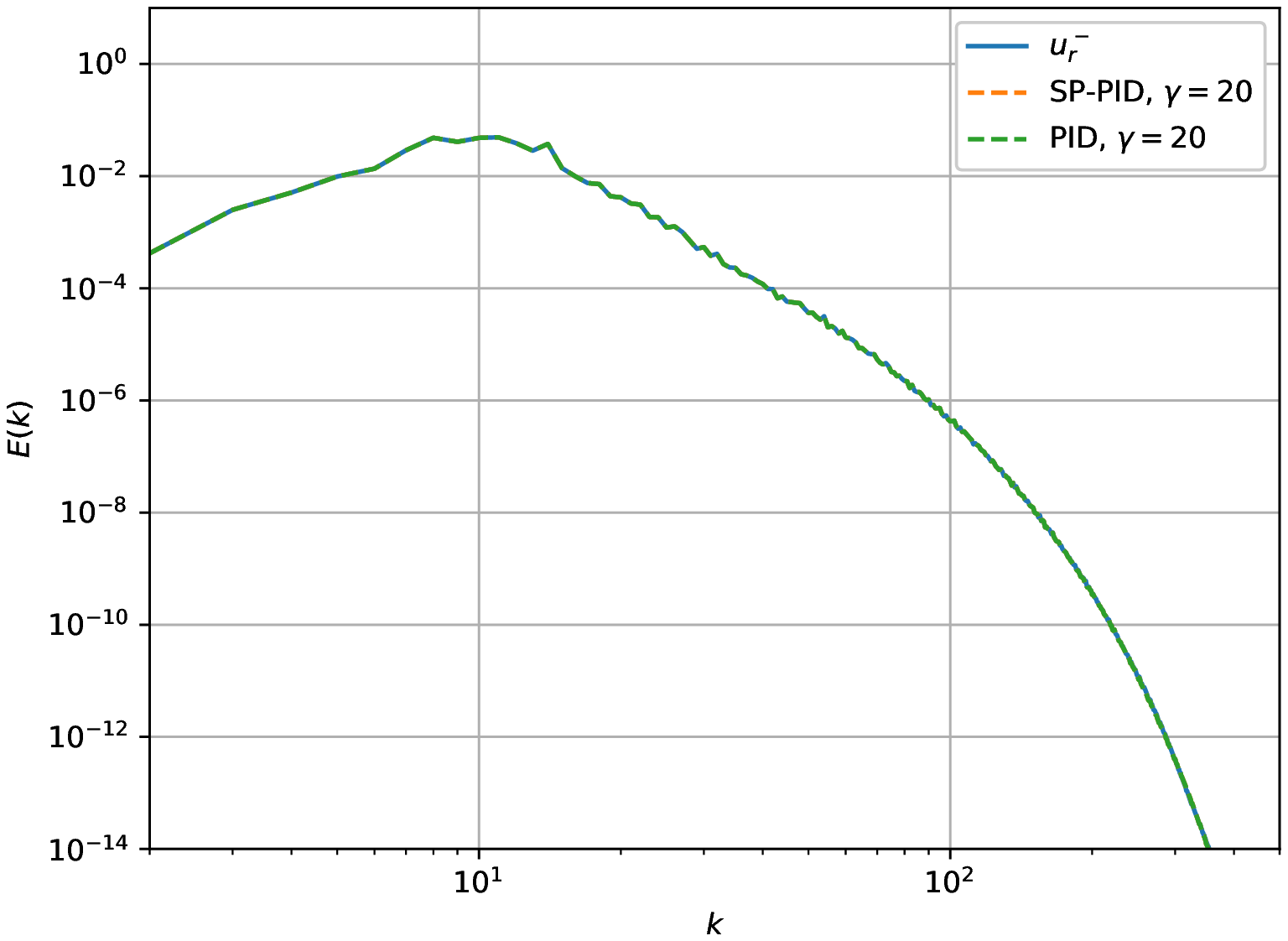} 
        \caption{Spectra before and after transition at $t = 0.5$ using PID and SP-PID for $\gamma = 20$.}
        \label{fig:umupom20}
    \end{minipage}
\end{figure}

\begin{figure}[h!]
    \centering
    \begin{minipage}{0.45\textwidth}
        \centering
        \includegraphics[width=1\textwidth]{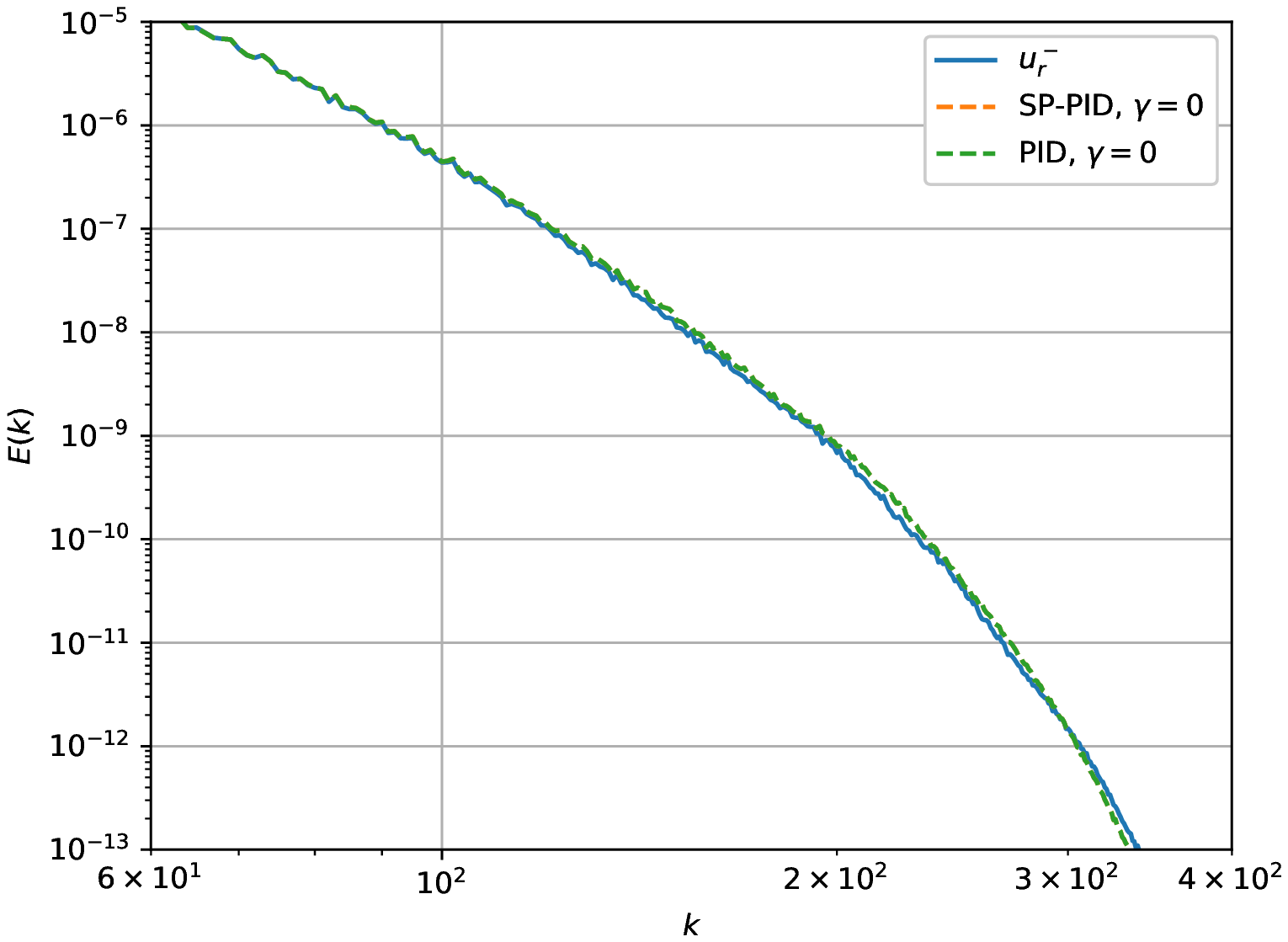} 
        \caption{Spectra before and after transition at $t = 0.5$ using PID and SP-PID for $\gamma = 0$ (zoomed).}
        \label{fig:umupom0z}
    \end{minipage}\hfill
    \begin{minipage}{0.45\textwidth}
        \centering
        \includegraphics[width=1\textwidth]{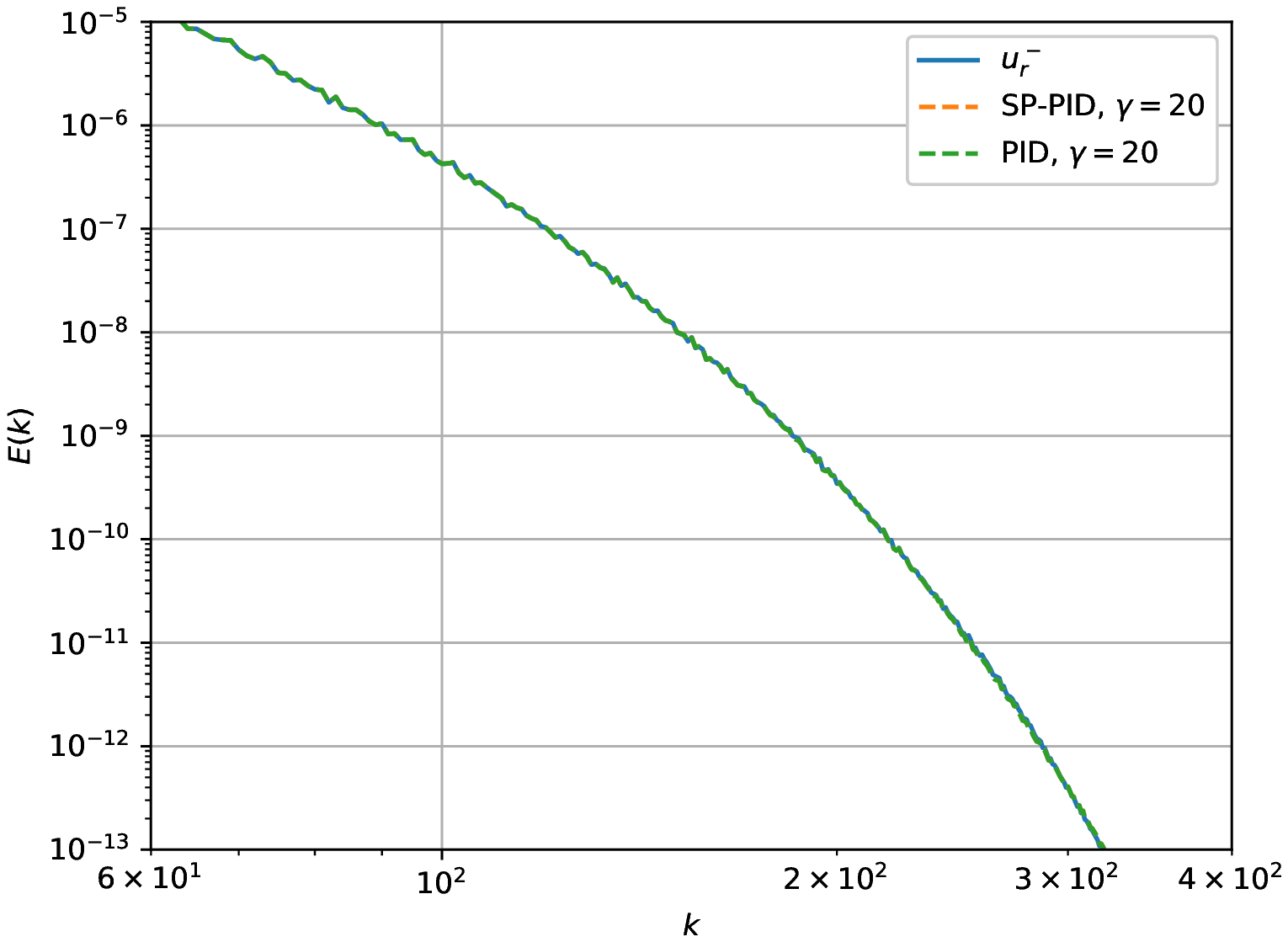} 
        \caption{Spectra before and after transition at $t = 0.5$ using PID and SP-PID for $\gamma = 20$ (zoomed).}
        \label{fig:umupom20z}
    \end{minipage}
\end{figure}

The results of the experiment have been displayed in \autoref{tab:errpid}. It can be seen that the transition accuracy indeed increases as the overlap parameter is increased, since the transition error $\epsilon_T$ decreases significantly. Furthermore, $\epsilon_T$ using the SP-PID and the PID is nearly identical for both overlap parameter values. This shows that even in the unconstrained setting the SP-PID interface condition \eqref{eq:intoptim} is nearly optimal. As expected, the conserved quantities are exactly conserved over interfaces using the SP-PID. Increasingly smaller errors are made using the standard PID as the overlap parameter increases. We have also shown the energy spectra of the velocity fields before ($\boldsymbol{u}_r^-$) and after ($\boldsymbol{u}_r^+$) the interface in \autoref{fig:umupom0} and \autoref{fig:umupom20} for $\gamma = 0$ and $\gamma = 20$, respectively. We see good agreement between the new and old spectra for both methods, however for larger overlap parameters the smaller scales in the flow are captured more accurately. This can be seen in the zoomed \autoref{fig:umupom0z} and \autoref{fig:umupom20z} for $\gamma = 0$ and $\gamma = 20$, respectively. From this we conclude that:
\begin{enumerate}
    \item the SP-PID exactly conserves both energy and momentum over interfaces, whereas the PID does not;
    \item the overall accuracy of the transition (conserved quantities and $\epsilon_T$) improves with greater overlap parameters $\gamma$ for the SP-PID and PID.
\end{enumerate}
We suspect there will be an optimal overlap parameter, this will be the subject of future research.

\section{Conclusion}
In this article we have proposed new structure-preserving hyper-reduction methods referred to as the CLSDEIM, OCLSDEIM and MCLSDEIM. In these new hyper-reduction methods the conventional DEIM is posed as a minimization problem and subsequently constrained to conserve kinetic energy, in addition to conserving momentum. The OCLSDEIM and MCLSDEIM enhance the robustness of the CLSDEIM by oversampling and by Mahalanobis regularization, respectively.

The second novelty in this work is that we have proposed a method to conserve energy and momentum over interfaces in order to use the new hyper-reduction methods in combination with the PID temporal localization method. The method is based on solving a constrained minimization problem at every interface, and its solution can be interpreted as a projection followed by a scaling to conserve energy. This new SP-PID allows the construction of fully structure-preserving and temporally localized hROMs of convection-dominated flows.

We have performed two numerical experiments to show the performance of the new structure-preserving hyper-reduction methods and the structure-preserving temporal localization method. We used the shear layer roll-up (SLR) to test the structure-preserving properties, accuracy and computational performance of the proposed hyper-reduction methods. We confirmed the theoretical results that the CLSDEIM, OCLSDEIM and MCLSDEIM conserve energy and momentum in a fully discrete setting when the Gauss-Legendre 4 (GL4) Runge-Kutta (RK) time integration method is used. It was also shown that the standard DEIM caused a significant error in kinetic energy in this case. Convergence experiments showed that both the DEIM and CLSDEIM exhibit erratic behaviour as a function of the DEIM space dimension $m$, due to error accumulation during the time integration process. This erratic behaviour and error accumulation is alleviated by the robust OCLSDEIM and MCLSDEIM. Finally, the computational performance of the DEIM, CLSDEIM and MCLSDEIM was comparable, whereas the OCLSDEIM was slower than the other methods. This shows the benefit of using regularization instead of oversampling to increase robustness of the CLSDEIM.

The second numerical experiment was the two-dimensional homogeneous isotropic turbulence (2DT). The results of this test case further highlighted the accuracy and robustness of the new hyper-reduction methods. Furthermore, the 2DT experiment showed that using the SP-PID energy and momentum can be conserved over interfaces, in contrast to the classical PID. Finally, it was shown that overlapping snapshots in the construction of local reduced spaces can increase the transition accuracy.

There are several directions of future research that we believe to be valuable or interesting. First of all is the generalization of the MCLSDEIM and SP-PID to parametric problems. For the MCLSDEIM this entails confirming that the DEIM coordinates of a parametric problem can indeed be thought of as originating from a single prior distribution. For the SP-PID this entails finding methods to efficiently and accurately distribute the full parametric snapshot set over multiple subsets. Besides these methods, the extension to arbitrary boundary conditions (instead of periodic conditions) will be very valuable. Finally, the methods proposed in this article could be merged with ideas from e.g.\ \cite{chaturantabutstructure, wangstructure} to obtain structure-preserving hROMs for problems with general first integrals \textit{and} general nonlinear skew-symmetric operators (like the convection operator in this article or those in \cite{miyatakestructure}). We believe this could be done by suitably altering condition \eqref{eq:deimcond} to conserve a general hyper-reduced first integral.

\section*{CRediT authorship contribution statement}
\textbf{R.B. Klein}: Conceptualization, Methodology, Software, Formal analysis, Investigation, Writing - Original Draft, Visualization \\
\textbf{B. Sanderse}: Conceptualization, Writing - Review \& Editing, Supervision, Project administration

\section*{Declaration of competing interests}
The authors declare that they have no known competing financial interests or personal relationships that could have appeared to influence the work reported in this paper.

\section*{Data availability}
Data will be made available on request.

\section*{Acknowledgements}
The authors thank R.A.W.M. Henkes for his supervision during the master thesis behind this work and thank C. Pagliantini for the stimulating discussions on discrete empirical interpolation during the MORe conference at TU Berlin and while visiting the CWI institute. This publication is part of the project ``Discretize first, reduce next'' (with project number VI.Vidi.193.105 of the research programme NWO Talent Programme Vidi which is (partly) financed by the Dutch Research
Council (NWO).



\bibliography{report}{}
\bibliographystyle{plain}

\end{document}